\documentclass[aps,prb,reprint,longbibliographystyle,superscriptaddress]{revtex4-2}
\usepackage{graphicx,dcolumn,bm,hyperref,amsmath,amssymb,subcaption,cleveref}
\usepackage{dcolumn}
\usepackage{bm}

\begin{document}
\title{Semiempirical \textit{ab initio} modeling of bound states of deep defects in semiconductor quantum technologies}
\author{YunHeng Chen}
\affiliation{Department of Quantum Science and Technology, Research School of Physics, Australian National University, Canberra, Australian Capital Territory 2601, Australia}
\author{Lachlan Oberg}
\affiliation{Department of Quantum Science and Technology, Research School of Physics, Australian National University, Canberra, Australian Capital Territory 2601, Australia}
\author{Johannes Flick}
\affiliation{Center for Computational Quantum Physics, Flatiron Institute, New York, NY 10010, U.S.A}
\affiliation{Department of Physics, CUNY-City College of New York, New York, NY 10031, U.S.A}
\affiliation{CUNY-Graduate Center, New York, NY 10016, U.S.A}
\author{Artur Lozovoi}
\affiliation{Department of Physics, CUNY-City College of New York, New York, NY 10031, U.S.A}
\author{Carlos A. Meriles}
\affiliation{Department of Physics, CUNY-City College of New York, New York, NY 10031, U.S.A}
\affiliation{CUNY-Graduate Center, New York, NY 10016, U.S.A}
\author{Marcus W. Doherty}
\affiliation{Department of Quantum Science and Technology, Research School of Physics, Australian National University, Canberra, Australian Capital Territory 2601, Australia}
\date{\today}

\begin{abstract}
A significant hurdle in developing high-performance semiconductor quantum technologies utilizing deep defects is related to charge dynamics. Unfortunately, progress in modeling their charge dynamics has been hindered over recent decades due to the absence of appropriate multiscale models capable of accurately representing the atomic properties of these defects and their impact on device performance. Here, we present a semi-\textit{ab initio} method for modeling the bound states of deep defects in semiconductor quantum technologies, applied to the negatively charged nitrogen vacancy (NV$^-$) center in diamond. We employ density functional theory calculations to construct accurate potentials for an effective mass model, which allow us to unveil the structure of the bound hole states. We develop a model to calculate the nonradiative capture cross sections, which agrees with experiment within one order of magnitude. Finally, we present our attempt at constructing the photoionization spectrum of NV$^0\rightarrow$ NV$^-$ + bound hole, showing that the electronic transitions of the bound holes can be distinguished from phonon sidebands. This paper offers a practical and efficient solution to a long-standing challenge in understanding the charge dynamics of deep defects.
\end{abstract}


\maketitle

\section{Introduction}
\par Deep defects are essential in semiconductor quantum technologies (SQT), finding diverse applications in quantum computing, sensing and communication \cite{Doherty2013,Awschalom2018,Wehner2018,Chen2020,Pezzagna2021,Nguyen2019,Bhaskar2020,Lohrmann2017,Son2020,Shaik2021,Naclerio2023}. Deep defects can exist in multiple charge states, posing a significant challenge for developing high-performance SQT due to the complexities of their charge dynamics \cite{Awschalom2018,Doherty2013,Lozovoi2023,Son2021,Gale2023}. Unfortunately, progress in modeling the charge dynamics of deep defects has stagnated over the past few decades due to the inherently complex multiscale problem, wherein atomic properties have a profound impact on macroscopic outcomes. Existing methods such as density functional theory (DFT) \cite{Jones2015} offer insights at the atomic scale but are limited by computational constraints, making them unsuitable for describing highly excited orbitals in larger volumes. On the other hand, phenomenological models lack the necessary atomic-level understanding to accurately predict properties such as capture cross sections and the photoionization spectrum. Overcoming these critical obstacles will significantly enhance the performance of SQT using deep defects.

\par In this paper, we overcome these deficiencies by presenting a semi-\textit{ab initio} model of the bound states of deep defects in semiconductor quantum technologies. For demonstration, we apply this model to the NV$^-$ center in diamond \cite{Doherty2013}. We first simulate the discretized energies of the bound hole states by adopting an effective mass Hamiltonian and use DFT to accurately describe the effective potential of NV$^-$. We then model the nonradiative capture of charge carriers by applying a deformation potential model of acoustic and optical hole-phonon scattering. Our results show that the simulated capture cross sections differ from experimental results by only one order of magnitude due to the exclusion of two-phonon Raman processes. Finally, we construct the photoionization spectrum of NV$^{0}$ $\rightarrow$ NV$^-$ + bound hole by evaluating the transition dipole moments using a combination of effective mass theory and DFT and by estimating the phonon sidebands (PSB). We show that the electronic transitions of the bound hole states are indeed distinguishable from the PSBs.

\section{Effective Mass Hamiltonian}
 In diamond, the valence band maximum (VBM) is triply degenerate \cite{Madelung2004}, where we approximate that the VBM consists of two degenerate heavy hole (HH) bands and a single light hole (LH) band. We simulate the energy levels of the bound hole states by adopting an effective mass Hamiltonian. This method assumes the solutions can be well approximated by an envelope wavefunction multiplied by a Bloch function \cite{Stoneham2001}. The coupling between each band is ignored for simplicity and to avoid ambiguities in the definition of the Bloch function basis when constructing the effective potential. The wavefunction solutions are given by
\begin{equation}
\psi_{b,i}\left(\vec{r}\right)=F_{b,i}\left(\vec{r}\right)u_b\left(\vec{r}\right)
\label{eq:wfsolution}
\end{equation}
where $u_{b}\left(\vec{r}\right)$ is the periodic Bloch function of the $b^{th}$ band at the valence band maximum (VBM) at the $\Gamma$ point of the unperturbed diamond structure, and $F_{b,i}\left(\vec{r}\right)$ is the slowly varying envelope wavefunction of the $b^{th}$ band of the $i^{th}$ level. This approximation is valid if the envelope wavefunction varies slowly within the primitive unit cell \cite{Stoneham2001}. 

\par Using the assumptions outlined above, we construct a Schr\"{o}dinger equation for the envelope wavefunction by taking the expectation value of the physical Hamiltonian with respect to the Bloch functions over the volume of a primitive unit cell and assigning an isotropic effective mass $(m_b)$ to each of the bands. Accordingly, the Schr\"{o}dinger equation is given by (see the Supplemental Material, SM \cite{boundholesupp})
\begin{equation}
\left(\frac{\vec{p}^2}{2m_b}+\Delta V_{b}\left(\vec{r}\right)\right)F_{b,i}\left(\vec{r}\right)=\Delta E_{b,i} F_{b,i}\left(\vec{r}\right)
\label{eq:effectivemassHamiltonian}
\end{equation}
where 
\begin{equation}
\Delta V_b\left(\vec{r}\right)=\frac{1}{V_c}\int_{-\infty}^{\infty} \left|f_{b}\left(\vec{r'}\right)\right|^2\Delta V\left(\vec{r'}+\vec{r}\right)d^3r'
\label{eq:beffectivepotential}
\end{equation}
The first and second term in Eq.~\ref{eq:effectivemassHamiltonian} are the kinetic energy operator with an effective hole mass tensor and the effective potential of the $b^{th}$ band, respectively. In Eq.~\ref{eq:beffectivepotential}, $f_b\left(\vec{r'}\right)$ is the Bloch function $u_b\left(\vec{r'}\right)$ within the central unit cell and is enforced to be zero outside of that volume of the unit cell, thereby allowing the integral limits to be expanded to all space. $\Delta V\left(\vec{r}\right)=V_{\text{NV$^-$}}\left(\vec{r}\right)-V_{\text{diamond}}\left(\vec{r}\right)$ is the effective potential defined by the difference between the potential generated by a NV$^-$ center ground state and a defect-free diamond. Note that $u_b\left(\vec{r}\right)$ are the solutions to a perfect diamond. $\Delta E_{b,i}=E_{b,i}-E_\text{VBM}$ is the difference in energy between the solution of the effective mass equation $(E_{b,i})$ and the recombination energy of NV$^0$ $(E_\text{VBM}=2.94~\text{eV})$ \cite{Aslam2013}.

\section{Constructing Effective Potential}
We adopt the frozen core approximation to define the effective potential. This assumes that the state of the electron at the NV center is not affected by the state of the hole, so the potential experienced by the hole is described by the electrostatic potential generated by the nuclei and electrons. However, this approximation explicitly ignores exchange-correlation interactions between the bound hole and the electrons in the NV center.

\par We construct the effective potential by employing DFT to obtain the electron charge densities and nuclear geometries of these two systems. The primitive unit cell of diamond consists of two atoms, and they contribute to the charge densities at the VBM. For demonstration purposes, we evaluate $\Delta V_b\left(\vec{r}\right)$ by approximating the probability density of VBM $\left|f_b\left(\vec{r'}\right)\right|^2$ as a single Gaussian function (see Sec. I B within the SM \cite{boundholesupp}).

\par After creating the near-field effective potential using DFT, the next step is to extrapolate it to larger volumes. While we anticipate the far-field potential to resemble that of a point charge due to the NV$^{-}$ center's effective single electron charge, there is also an intermediate-field potential that deviates from this behavior. To capture this potential, we solve the Poisson's equation in COMSOL Multiphysics to obtain the intermediate-field potential. 
\begin{figure}[h!]
\centering
	\begin{subfigure}[t]{0.5\textwidth}
	\includegraphics[width=0.95\textwidth]{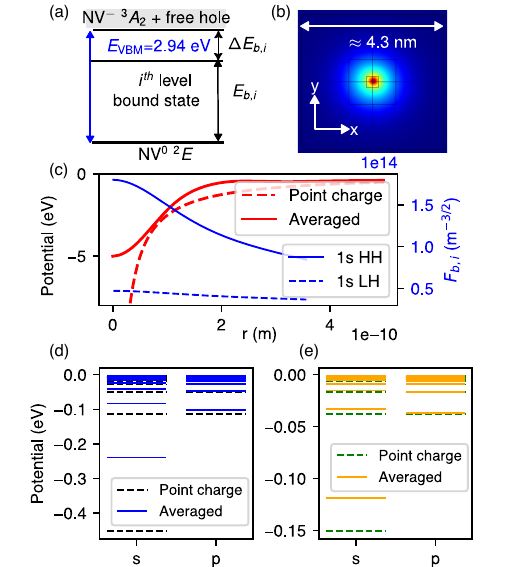}
	\phantomcaption
	\label{fig:1a}
	\end{subfigure}
	\begin{subfigure}[t]{0\textwidth}
	\includegraphics[width=\textwidth]{boundstates}
	\phantomcaption
	\label{fig:1b}
	\end{subfigure}
		\begin{subfigure}[t]{0\textwidth}
	\includegraphics[width=\textwidth]{boundstates}
	\phantomcaption
	\label{fig:1c}
	\end{subfigure}
		\begin{subfigure}[t]{0\textwidth}
	\includegraphics[width=\textwidth]{boundstates}
	\phantomcaption
	\label{fig:1d}
	\end{subfigure}
		\begin{subfigure}[t]{0\textwidth}
	\includegraphics[width=\textwidth]{boundstates}
	\phantomcaption
	\label{fig:1e}
	\end{subfigure}
\caption{Simulation of bound hole states. (a) Schematic description of the bound hole states. The shaded region denotes the continuum levels of valence band states. (b) Wavefunction of the ground bound hole state simulated using COMSOL. (c) Comparison of a spherically averaged potential and a point charge potential in diamond and the $l=0$ radial envelope wavefunctions for HH and LH. The radial wavefunctions are plotted over the dimension of a primitive unit cell with a radius of $3.567~\text{\AA}$. (d) Comparison of the $l=0$ and 1 bound HH and (e) LH states (solid lines) with a hydrogenic model (dashed lines) using the stated effective masses.}
\label{fig:1}
\end{figure}
\section{Simulation of Bound Hole States}
We solve Eq.~\ref{eq:effectivemassHamiltonian} using three different methods. Firstly, we use COMSOL simulations with a Dirichlet boundary condition and an isotropic effective mass to obtain the low lying excited states (ES). However, directly obtaining the higher excited bound hole states via COMSOL simulations is computationally inefficient due to the large spatial extent of their wavefunctions. Thus, we use a spherically averaged effective potential to solve for the radial Schr\"odinger equation to obtain the higher ES. Lastly, we benchmark the solutions of the first two methods using a simple hydrogenic model.
\par Fig.~\ref{fig:1b} shows the ground state (GS) energy due to a HH, obtained from COMSOL simulations. The simulation used isotropic effective masses of 1.08$m_e$ and 0.36$m_e$ for the heavy and light hole mass, respectively (where $m_e$ is the mass of the electrons; see the SM \cite{boundholesupp}) over a dimension of $214.5\times 214.5\times 214.5~\text{\AA}^3$. The GS energy is -0.24~eV below $E_\text{VBM}$. We achieve mesh convergence with deviations of less than 1\% for both HH and LH solutions and cell convergence with errors of less than 7\%(3\%) up to 30(5) levels for HH(LH) solutions. Using $E_\text{VBM}=2.94~\text{eV}$ \cite{Aslam2013}, we find the GS simulated using our methods (-0.24~eV) to be in good agreement with the energy obtained via \textit{ab initio} calculations (-0.27~eV) \cite{Lozovoi2021}. As the exchange-correlation potential is not captured in our EMA model, this result sets an upper limit of 0.03~eV on the effects of exchange correlation, which decrease as the hole orbit gets larger.  
 
\par Comparing the spherically averaged solutions to COMSOL, COMSOL observes additional splittings due to the true $C_{3v}$ symmetry of the defect, unlike the former. However, HH and LH solutions between these methods vary by less than 3\% when ignoring these splittings and considering mean values of energy levels (see the SM \cite{boundholesupp}). This agreement prompts us to examine the spherically averaged solutions, revealing a hydrogenic-like potential at large distances. This implies that higher ES solutions will be well approximated by a hydrogenic series [Fig.~\ref{fig:1c}]. Comparing eigenenergies of HH and LH with a hydrogenic model, the higher ES with azimuthal quantum number $l=0$ and 1 indeed closely follow a hydrogenic series, unlike the lower ES [Figs.~\ref{fig:1d} and \ref{fig:1e}]. Our simulations confirm the validity of using EMA since the radial wavefunctions exhibit minimal changes within the primitive unit cell, except for the GS where solution accuracy may diminish. [Fig.~\ref{fig:1c}]. Moreover, this method is efficient and enables us to access higher excited bound hole states that are otherwise unattainable in three-dimensional simulations.

\begin{figure}[h!]
\centering
\includegraphics[width=0.5\textwidth]{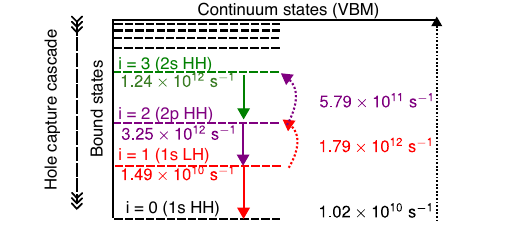}
\caption{Schematics for hole cascade capture process via emission of phonons. Solid and dotted lines, along with their respective values denote the emission rates and absorption rates, respectively.}
\label{fig:holecapture}
\end{figure}
\section{Capture Cross Section}
We now model the nonradiative capture of the holes by applying the deformation potential model of hole-acoustic/optical phonon scattering (see Sec.~V within the SM \cite{boundholesupp}). The hole capture cross section can be determined rigorously using the cascade trapping framework, mediated by the emission of phonons \cite{Lax1960,Hamann1964,Stoneham2001}. The capture process occurs when phonon emission is much faster than absorption, such that the bound hole has a negligible chance of excitation. The critical transition occurs from the first ES to the GS. As we will be showing later, we consider this transition to be the capture event as the absorption of phonons out from the GS is slower than the emission of phonons to the GS from the first ES. Our previous paper which utilized a brute-force Monte Carlo simulations \cite{Lozovoi2023} demonstrated that multiple inelastic scattering events helped the system to reach the critical transition at room temperature. Consequently, the justification for this hole capture model, contingent upon the ability of a hole to cascade down to this critical transition, is firmly established.

\par We calculate the scattering rates for all possible direct transitions from and to the GS, first ES and the $n$s HH states $(n=2-5)$ (see the SM \cite{boundholesupp}). We find that the scattering rates are decreasing with increasing $n$. However, the asymptotic limit of the rates is unclear due to limited datasets (see the SM \cite{boundholesupp}). We observe that the direct emission rate from the first ES to the GS is faster compared to the total absorption rate out from the GS (Fig.~\ref{fig:holecapture}). Hence, the emission rate of this critical transition is effectively the hole capture rate. As the emission rate of the first ES is slower than the absorption rate to the second ES, we can then assume that thermal equilibrium (TE) has been established within the ES before it makes the critical transition down to the GS. Subsequently, the capture rate is approximated as the sum of the total emission rate from all the ES to the GS, multiplied by the partition function.

\par The capture cross section $\sigma_{\text{cap}}$ can be defined as \cite{Alkauskas2014}
\begin{equation}
\sigma_\text{cap}  = \frac{V \Gamma_\text{cap}}{\nu}
\end{equation}
where $\nu$ is the hole thermal velocity in diamond, $V$ is the volume of the system, and $\Gamma_\text{cap}$ is the hole capture rate. Assuming the TE approximation, both bound and free hole states are considered in the entire diamond, with the latter being dominant. Hence, the total partition function $Z_{\text{tot}}$ is the sum of the partition functions due to the bound states $Z_\text{B}$ and the free hole states $Z_\text{F}$. 

 \par By converting the discrete sum of wavevectors into a continuous variable, $Z_\text{F}$ is proportional to $V$. We adopt the partition function of a hydrogen atom for the bound state \cite{Ebeling2012} and we terminate the sum over the bound states by the maximum principal quantum number that corresponds to the wavefunctions, which fits within $V$, determined by the Bohr's radius of the wavefunction. Subsequently, $Z_\text{B}$ is proportional to $V^{1/3}$. In the limit of large $V$, the contribution of $Z_\text{B}$ is negligible compared to $Z_\text{F}$ and thus the capture cross section, valid at room temperature is approximately (see the SM \cite{boundholesupp})
\begin{equation}
\sigma_\text{cap}\approx \frac{8}{\nu}e^{-\beta E_1}\left(\pi\kappa\beta\right)^{3/2}  \sum_{b,i_\text{B}}e^{-\beta\Delta E_{b,i_\text{B}}}\Gamma_0^{b,i_\text{B}}
\label{eq:capturecs}
\end{equation}
where $E_1$ is the energy of the first ES, $\beta=1/k_BT$ in which $k_B$ is the Boltzmann constant and $T$ is the temperature. $\kappa$ is the curvature averaged over the HH and LH bands at the VBM. $\Delta E_{b,i_\text{B}}$ is the energy difference between the $i^{th}$ bound state and $E_1$ and $\Gamma_0^{b,i_\text{B}}$ denotes all possible direct emission rates from the higher bound states to the GS permitted by the phonon energies. Eq.~\ref{eq:capturecs} is a physically valid expression that doesn't depend on the volume of the system.

\par The capture cross section at $T=300$~K is $1.65\times 10^{-4}\mu\text{m}^{2}$, approximately one order of magnitude smaller than the experimental value \cite{Lozovoi2021}. One-phonon process dominates at low temperatures \cite{Astner2018} whereas two-phonon processes dominate at room temperature \cite{Doherty2013}. The discrepancy in our calculations is attributed to the exclusion of two-phonon processes in the scattering rate calculations. Two-phonon emission rates will be larger due to the additional decay pathways accessible through the combination of a thermally driven and a spontaneous transition. However, this is not the case for absorption processes which necessitate two thermally driven transitions. At room temperature, such thermal processes are significantly slower than the dominant spontaneous emission transitions, as the relevant phonon energy is much larger than the thermal energy. Thus, the capture rate is expected to increase due to higher overall emission rate. Consequently, we underestimated the calculated capture cross sections. To prevent divergence at the low-temperature limit, we will need to include the contributions from the bound states to Eq.~\ref{eq:capturecs}, which is now volume dependent.

\section{Photoionization spectrum of NV$^0$ $\rightarrow$ NV$^-$ + bound holes}
We now turn to modeling the photoionization spectrum of NV$^0$ $\rightarrow$ NV$^-$ + bound holes. It is impossible to directly model this using \textit{ab initio} techniques due to computational constraints. Instead, we make an approximation by adopting the Huang-Rhys model to describe the zero-phonon lines (ZPL) and the PSBs \cite{Stoneham2001}. The spectrum can be obtained by evaluating the transition dipole moment, constructing the vibrational overlap functions and subsequently evaluating the absorption cross section (see the SM \cite{boundholesupp}). The bound-state solutions are represented by Eq.~\ref{eq:wfsolution}. The observed transition is the transition of a hole from the $e$ orbital of NV$^0$ to the bound state. Here, we assume that the $e$ orbital of NV$^0$ is sufficiently localized that it is only nonzero over the dimension of the NV$^-$ center. In this region, it is suitable to perform Taylor series expansion on the dipole moment transition matrix elements $\left(\vec{D}_{b,i}\right)$ in terms of the envelope wavefunctions obtained from the bound-state simulations and the integrals over $e$ orbital and $u_{b}\left(\vec{r}\right)$, which are evaluated using DFT. In this paper, we only deal with the zero-order term and thus limit ourselves to transitions involving $s$-like orbitals. 
\par The presence of an NV center reduces the cubic symmetry of diamond to $C_{3v}$ symmetry and splits the triply degenerate states at $\Gamma$ point of the VBM to a pair of doubly degenerate states and one single degenerate state. Using DFT, we identified two degenerate transitions originating from the valence band states to the $e_y$ orbital of NV$^0$, which corresponds to HH. However, LH transition could not be identified. We then averaged the absolute square of the transition dipole moments from the doubly degenerate valence band states, yielding a parallel-to-transverse component ratio of 0.32. Consequently, the absorption cross section is polarization independent (see the SM \cite{boundholesupp}).

\par Taking into account the Stokes shift $(\Delta E_{S})$, the absorption cross section consisting of ZPL and PSBs, summed over all possible transitions at energy $E_{b,i}$ is given by
\begin{eqnarray}
\sigma\left(E\right)&=& 4\pi^2\chi \sum_{b,i}E \overline{\left|\vec{D}_{b,i}\right|^2} \Bigg(L\left(E-E_{b,i}+\Delta E_S,\Gamma_{b,i}\right)\nonumber \\ 
&&+P\left(E-E_{b,i}+\Delta E_S,T\right)\Bigg)
\end{eqnarray}
 where $\chi=e^2/4\pi\epsilon\hbar c$ is the fine structure constant of diamond, $L$ is ZPL of each transition, $P$ is the vibrational overlap functions and $\overline{\left|\vec{D}_{b,i}\right|^2}$ is the averaged transition dipole matrix elements. Both features are explicitly weighted by the exponent of the temperature-dependent Huang-Rhys factor $\left(S\right)$. 

\par
\begin{figure}[h]
\centering
\includegraphics[width=0.5\textwidth]{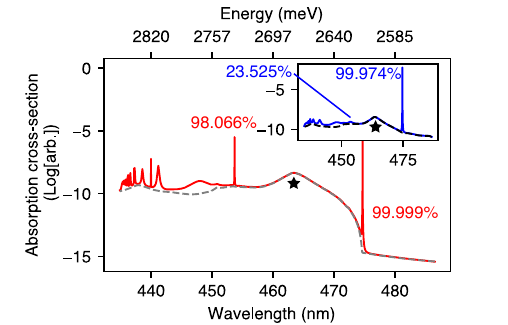}
\caption{Absorption cross section at $T = 5~$K plotted on a logarithmic scale and renormalized relative to the highest peak of the spectrum. The solid lines represent the full photoionization spectrum whereas the dotted lines are the PSBs. The PSBs for wavelength greater than 475~nm at $T = 5~$K have been reconstructed to prevent divergence on the logarithmic scale. The star signs denote the phonon sideband peak at 463~nm. The percentages indicate the fidelities of exciting a ZPL peak, labeled for the first and second bound hole transitions. (Inset) Absorption cross section at $T = 300~$K.}
\label{fig:absorptioncrosssection}
\end{figure}

 We make the following assumptions to obtain the absorption cross sections. First, we validated the similarities of the phonon modes of the internal transitions of NV$^-$ and $\text{NV}^0 \rightarrow \text{NV}^-+h^+$ using DFT, which agree to 71\%. Consequently, we can construct the PSBs of the NV$^0$ $\rightarrow$ NV$^{-}$ + bound holes by using the known generating function of the NV$^-$ PSBs \cite{Kehayias2013}, combined with an estimate of the Huang-Rhys factor and the evaluated energies of those transitions. Second, using DFT and conventional methods, we obtained the low-temperature Huang-Rhys factor of $S_0=0.942$ for the NV$^0$ $\rightarrow$ NV$^-$ + $h^+$ (free hole) transition with the assumption of $C_{3v}$ symmetry and have explicitly neglected the Jahn-Teller effect. We adopted the Huang-Rhys factor for the transitions involving the bound states. Finally, we modified $S$ \cite{Goldman2015} and constructed the PSBs accordingly at $T=5~$K and $T=300~$K, as depicted in Fig.~\ref{fig:absorptioncrosssection}.

 \par We model the ZPL of each transition with a Lorentzian function. The Lorentzian function has a linewidth defined by the total scattering rate $\left(\Gamma_{b,i}\right)$ from each transition due to the absorption and emission rate as obtained from the hole-acoustic/optical phonon scattering model (see the SM \cite{boundholesupp}). We have explicitly neglected the recombination rate as we expect the recombination to be much slower than other broadening processes. For energy levels that have not been explicitly calculated, we assume that their linewidths exhibit a scattering rate that is similar to that of the preceding energy level. Additionally, we approximate the GS scattering rate at low temperature to be limited by the lifetime of the ES, which is typically on the order of 10~ns \cite{Doherty2013}.

\par The electronic transitions of the bound states can be distinguished from the PSBs as seen in Fig.~\ref{fig:absorptioncrosssection}. However, it is crucial to note that there is a PSB peak at 463 nm, which may be misidentified as a ZPL. For Rydberg-like schemes, the ability to selectively excite a specific electronic transition is crucial. The key measure is the fidelity of exciting a ZPL peak compared to the phonon sideband beneath it. The calculated high selectivity at low temperatures is promising for the potential application of Rydberg-like schemes.
\section{Conclusion}
\par In summary, we presented a semi-\textit{ab initio} model for simulating the bound states of deep defects in semiconductor quantum technologies, which we applied to the NV$^-$ center in diamond. We quantitatively showed the model's prediction of the lowest bound-state energy to be in good agreement with direct \textit{ab initio} simulations. We used the solutions of the bound states to model hole capture cross section, which agrees with experiments within one order of magnitude and to construct the first prediction of photoionization spectrum of NV$^0\rightarrow$ NV$^-$ + bound hole. These models represent a critical advancement to a long-standing challenge in understanding deep defects’ charge dynamics. However, additional work is admittedly in order. This includes a better description of the probability density of VBM charge densities, incorporating two-phonon Raman processes in the scattering rate calculations, and accurate modeling of NV$^0$ PSBs. 

\acknowledgments{A.L. and C.A.M acknowledge support from the National Science Foundation through Grants No.~NSF-2216838 and NSF-1914945; they also acknowledge access to the facilities and research infrastructure of the NSF CREST IDEALS, Grant No.~NSF-2112550. M.W.D. acknowledges support from the Australian Research Council DE170100169. The Flatiron Institute is a division of the Simons Foundation.}


\bibliography{ms.bbl}
\end{document}


\title{Supplemental Material for ``Semi-empirical \textit{ab initio} modeling of bound states of deep defects in semiconductor quantum technologies''}
\author{YunHeng Chen}
\affiliation{Department of Quantum Science and Technology, Research School of Physics, Australian National University, Canberra, Australian Capital Territory 2601, Australia}
\author{Lachlan Oberg}
\affiliation{Department of Quantum Science and Technology, Research School of Physics, Australian National University, Canberra, Australian Capital Territory 2601, Australia}
\author{Johannes Flick}
\affiliation{Center for Computational Quantum Physics, Flatiron Institute, New York, NY 10010, U.S.A}
\affiliation{Department of Physics, CUNY-City College of New York, New York, NY 10031, U.S.A}
\affiliation{CUNY-Graduate Center, New York, NY 10016, U.S.A}
\author{Artur Lozovoi}
\affiliation{Department of Physics, CUNY-City College of New York, New York, NY 10031, U.S.A}
\author{Carlos A. Meriles}
\affiliation{Department of Physics, CUNY-City College of New York, New York, NY 10031, U.S.A}
\affiliation{CUNY-Graduate Center, New York, NY 10016, U.S.A}
\author{Marcus W. Doherty}
\affiliation{Department of Quantum Science and Technology, Research School of Physics, Australian National University, Canberra, Australian Capital Territory 2601, Australia}
\date{\today}


\maketitle


\section{\textit{Ab initio} calculations of the charge densities}
\par Density functional theory (DFT) was performed using the VASP plane-wave code \cite{Kresse1996a,Kresse1994,Kresse1996,Kresse1999} using PBE\cite{Perdew1996} exchange-correlation functionals for accurate determination of the ground state (GS) geometry and electronic properties. We employ a 512-atom cubic unit cell, $8\times8\times8$ Monkhorst-Pack $k$-point sampling, and a plane wave cutoff of 900~eV. The electronic solutions are converged to a tolerance of $10^{-5}$~eV and the ionic geometry is optimized to a force tolerance of $10^{-4}$~eV/\text{\AA} per ion. From these calculations, we extract the total charge density of a defect-free diamond and NV$^-$ GS, and charge densities of the Kohn-Sham orbitals at the diamond valence band maximum (VBM).
\subsection{Charge densities of NV$^-$ and a defect-free diamond}
\label{sec:chargedensityNV}
\begin{figure}[h!]
\centering
    \begin{subfigure}[t]{0.95\textwidth}
    \includegraphics[width=\textwidth]{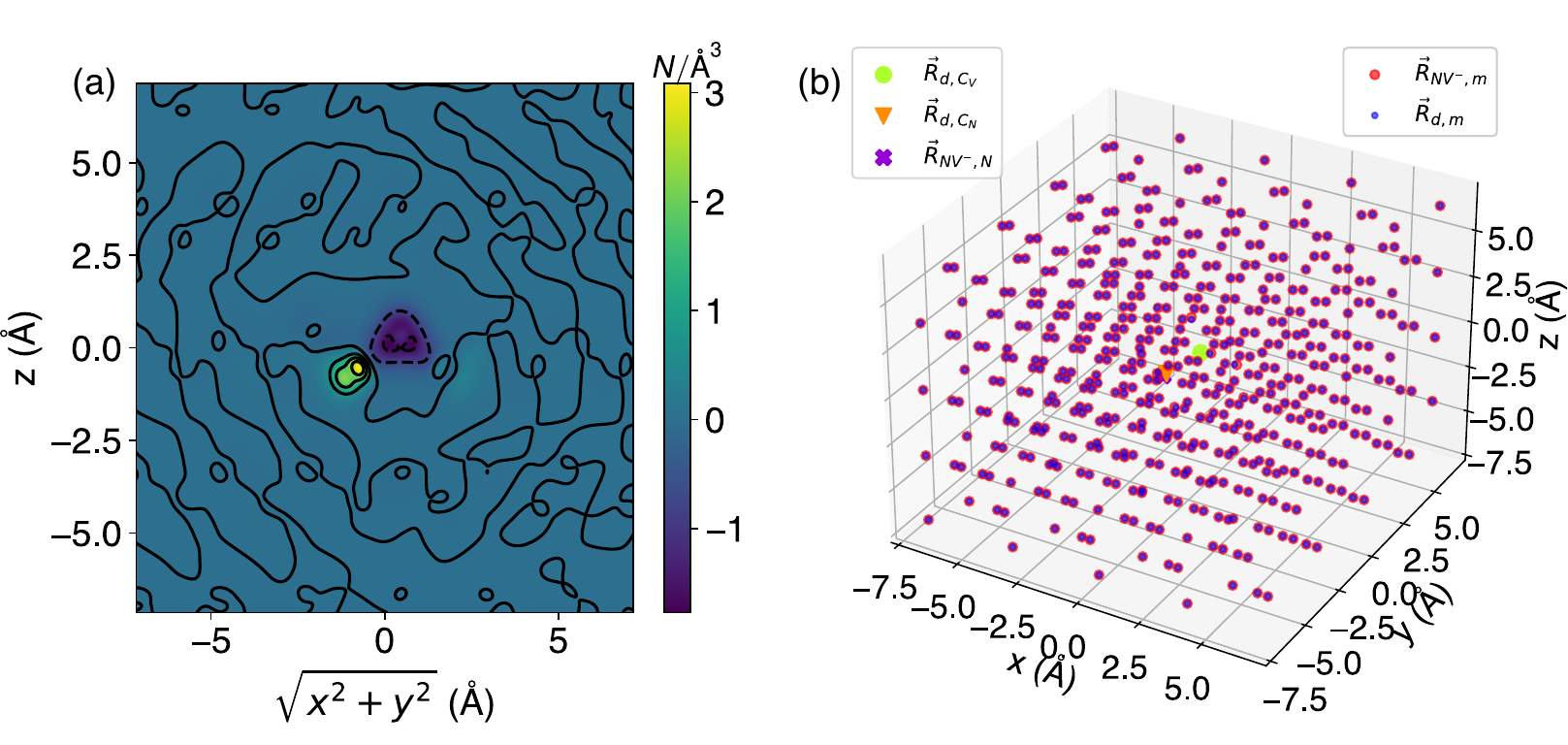}
    \phantomcaption
    \label{fig:chgdensity}
    \end{subfigure}
    \begin{subfigure}[t]{0\textwidth}
    \includegraphics[width=\textwidth]{chgdiff_coordinates.pdf}
    \phantomcaption
    \label{fig:coordinate}
    \end{subfigure}
\caption{(a) Difference in valence band charge densities between an NV$^-$ center and a defect-free diamond. The charge densities are shifted towards the center of charge distributions and plotted over the (110) plane. The yellow and purple spots denote the location of the N atom (positive difference) and the vacancy (negative difference), respectively. (b) The shifted nuclear charge coordinates. $\vec{R}_{\text{NV}^-,m}$  and $\vec{R}_{\text{d},m}$ are the positions of the $m^{th}$ carbon in the NV and perfect diamond, respectively. $\vec{R}_{\text{NV}^-,\text{N}}$ is the position of the nitrogen atom in the NV, $\vec{R}_{\text{d},\text{C}_\text{N}}$ and $\vec{R}_{\text{d},\text{C}_\text{V}}$ are the positions of the carbon atoms at the nitrogen and vacancy sites, respectively, in a perfect diamond.}
\label{fig:chgdensity_coordinate}
\end{figure}
\par From the \textit{ab initio} calculations, the units cells have a dimension of $14.3\times 14.3\times 14.3~\text{\AA}^3$. The center of charge distribution $\vec{r}_\text{center}$ is given by
\begin{equation}
\vec{r}_\text{center}=\frac{1}{Q_\text{net}}\int_{0}^{r=14.3~\text{\AA}} \vec{r}\left(\rho_\text{NV$^-$}\left(\vec{r}\right) -\rho_\text{diamond}\left(\vec{r}\right)\right) d^3r
\end{equation}
where $Q_\text{net}=-2$ is the total number of valence electrons in the volume, $\rho_\text{NV$^-$}\left(\vec{r}\right)$ and $\rho_\text{diamond}\left(\vec{r}\right)$ are the valence band charge densities of NV$^-$ and a defect-free diamond, respectively. We obtain $\vec{r}_\text{center}=\left(4.156,4.156,4.156\right)$ \AA.
\par We then shift the coordinates of the NV$^-$ and diamond valence band charges and nuclear charge coordinates with respective to $\vec{r}_\text{center}$, resulting in Fig.~\ref{fig:chgdensity} and \ref{fig:coordinate}. Fig.~\ref{fig:chgdensity} shows the difference in charge densities between these two systems, plotted over the (110) plane whereas the resulting nuclear charge coordinates are shown in the Fig.~\ref{fig:coordinate}. We obtain $\vec{R}_{\text{NV}^-,\text{N}}=\left(-0.664,-0.664,-0..664\right)$~\AA, $\vec{R}_{\text{d},\text{C}_\text{N}}=\left(-0.579,-0.579,-0.579\right)$~\AA ~and $\vec{R}_{\text{d},\text{C}_\text{V}}=\left(0.314,0.314,0.314\right)$~\AA

\subsection{Valence band maximum charge densities of a bulk diamond}
\begin{figure}[h!]
\centering
\includegraphics[width=0.5\textwidth]{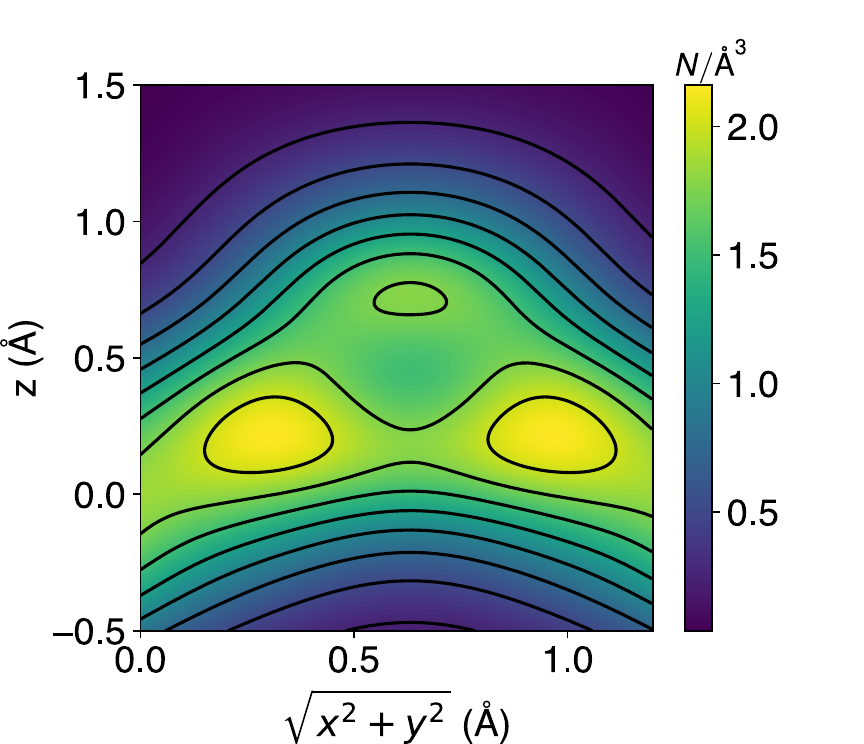}
\caption{The sum of charge densities of the spin up channel of the VBM states of a bulk diamond. The charge densities are plotted over the dimension of a primitive unit cell consisting of two atoms.}
\label{fig:vbmchg}
\end{figure}

\par Fig.~\ref{fig:vbmchg} shows the charge densities of the sum of three VBM states (spin up channel) of a bulk diamond. We observe that the primitive unit cell consists of two blobs of charge densities. For the purpose of demonstration, we will approximate these charge densities as a single Gaussian function (see Sec.~\ref{sec:simplestEM}). A single Gaussian description of the true VBM charge densities is a significant simplification. Future works include a more sophisticated fit of these charge densities or a atomic orbitals description.
\section{Effective mass model of bound holes of NV$^-$}
\subsection{Derivation of an effective mass model of bound holes of NV$^-$}
In the single particle picture, the states $\psi$ occupied by holes are solutions of the Schr\"odinger equation
\begin{equation}
\left(T+V_D+\Delta V\right)\psi=E\psi 
\label{eq:SE}
\end{equation}
where $T$ is the kinetic energy operator, $V_d$ is the unperturbed crystal potential and $\Delta V$ is the perturbing potential resulting from the introduction of the NV$^-$ center. If we consider only states whose energies are close to the VBM, then the following ansatz is expected to be a good solution
\begin{equation}
\psi\left(\vec{r}\right)=\sum_b F_b\left(\vec{r}\right)u_b\left(\vec{r}\right)
\label{eq:wf}
\end{equation}
where $u_b\left(\vec{r}\right)$ is the periodic Bloch function of the $b^{th}$ degenerate band at the VBM at the $\Gamma$-point of the unperturbed diamond structure, and $F_b\left(\vec{r}\right)$ is the unknown envelope function of the $b^{th}$ band. In diamond, there are three degenerate bands at the VBM \cite{Madelung2004}. We approximate them to be two heavy hole bands (HH), and one light hole band (LH). Substituting this ansatz into the Schr\"odinger equation above and applying the product rule to the kinetic energy operator, we find
\begin{equation}
\sum_bu_b\left(\vec{r}\right)TF_b\left(\vec{r}\right)+\Delta VF_b\left(\vec{r}\right)u_b\left(\vec{r}\right)+\alpha \vec{\nabla}F_b\left(\vec{r}\right)\cdot \vec{\nabla}u_b\left(\vec{r}\right)=\Delta E\sum_b F_b\left(\vec{r}\right)u_b\left(\vec{r}\right)
\label{eq:SEexpand1}
\end{equation}
where $\Delta E=E-E_\text{VBM}$, $E_\text{VBM}$ is the valence band maximum energy, and $\alpha =-\hbar^2/2m_e$. $\Delta E$ arises as we know the terms where the envelope function is multiplying the energy of a hole in a perfect diamond. Subsequently, left multiplying by $u^*_{b'}\left(\vec{r}\right)$ and integrating over the volume of the primitive unit cell at position $\vec{R}_\alpha$, we find
\begin{eqnarray}
&&\sum_b\int_{V_\alpha}u^*_{b'}\left(\vec{r}\right)u_b\left(\vec{r}\right)TF_b\left(\vec{r}\right)d^3r
+\int_{V_\alpha}u^*_{b'}\left(\vec{r}\right)u_b\left(\vec{r}\right)\Delta V\left(\vec{r}\right)F_b\left(\vec{r}\right)d^3r
+\alpha\int_{V_\alpha}u^*_{b'}\left(\vec{r}\right)\vec{\nabla}u_b\left(\vec{r}\right)\cdot\vec{\nabla}F_b\left(\vec{r}\right)d^3r\nonumber \\ 
&&=\Delta E\sum_b\int_{V_\alpha}F_b\left(\vec{r}\right)u^*_{b'}\left(\vec{r}\right) u_b\left(\vec{r}\right)d^3r
\label{eq:SEexpand2}
\end{eqnarray}
 Now, over the dimensions of a primitive unit cell, the envelope functions are expected to be approximately constant, thus making this approximation, the equation above becomes 
\begin{eqnarray}
&&\sum_b\int_{V_\alpha}u^*_{b'}\left(\vec{r}\right)u_b\left(\vec{r}\right)d^3rTF_b\left(\vec{r}\right)\big|_{\vec{R}_\alpha}
+\int_{V_\alpha}u^*_{b'}\left(\vec{r}\right)u_b\left(\vec{r}\right)\Delta V\left(\vec{r}\right)d^3rF_b\left(\vec{R}_\alpha\right)
+\alpha\int_{V_\alpha}u^*_{b'}\left(\vec{r}\right)\vec{\nabla}u_b\left(\vec{r}\right)d^3r\cdot\vec{\nabla}F_{b}\left(\vec{r}\right)\big|_{\vec{R}_\alpha}\nonumber \\ 
&&=\Delta E\sum_bF_b\left(\vec{R}_\alpha\right)\int_{V_\alpha}u^*_{b'}\left(\vec{r}\right) u_b\left(\vec{r}\right)d^3r
\label{eq:SEexpand3}
\end{eqnarray}
Using the following orthonormality condition for the Bloch functions
\begin{equation}
\int_{V_\alpha}u_{b'}^*\left(\vec{r}\right)u_b\left(\vec{r}\right)d^3r=V_c\delta_{bb'} 
\end{equation}
where $V_c$ is the volume of the primitive unit cell, then Eq.~\ref{eq:SEexpand3} becomes

\begin{eqnarray}
&&\sum_b\delta_{bb'}TF_b\left(\vec{r}\right)\big|_{\vec{R}_\alpha}
+\frac{1}{V_c}\int_{V_\alpha}u^*_{b'}\left(\vec{r}\right)u_b\left(\vec{r}\right)\Delta V\left(\vec{r}\right)d^3rF_b\left(\vec{R}_\alpha\right)
+\frac{\alpha}{V_c}\int_{V_\alpha}u^*_{b'}\left(\vec{r}\right)\vec{\nabla}u_b\left(\vec{r}\right)d^3r\cdot\vec{\nabla}F_b\left(\vec{r}\right)\big|_{\vec{R}_\alpha}\nonumber \\ 
&&=\Delta E\sum_b\delta_{bb'}F_b\left(\vec{R}_\alpha\right)
\end{eqnarray}
This can be rewritten in tensor form as
\begin{equation}
\overleftrightarrow{T}\cdot\vec{F}\left(\vec{r}\right)\big|_{\vec{R}_\alpha}+\overleftrightarrow{\Delta V}\left(\vec{R}_\alpha\right)\cdot\vec{F}\left(\vec{R}_\alpha\right)+\alpha\overleftrightarrow{P}\cdot\vec{F}\left(\vec{r}\right)\big|_{\vec{R}_\alpha}=\Delta E\vec{F}\left(\vec{R}_\alpha\right)
\end{equation}
where $\vec{F}$ is the vector of envelope functions for different Bloch functions, $\overleftrightarrow{T}=T\overleftrightarrow{I}$, $\overleftrightarrow{I}$ is the identity tensor, and the $(b,b')$ elements of the other tensors are defined as
\begin{eqnarray}
\overleftrightarrow{\Delta V}_{bb'}\left(\vec{R}_\alpha\right)&=&\frac{1}{V_c}\int_{V_\alpha}u^*_{b'}\left(\vec{r}\right)u_b\left(\vec{r}\right)\Delta V\left(\vec{r}\right)d^3r\\
\overleftrightarrow{P}_{bb'}&=&\frac{1}{V_c}\int_{V_\alpha}u^*_{b'}\left(\vec{r}\right)\vec{\nabla}u_b\left(\vec{r}\right)d^3r\cdot\vec{\nabla}
\end{eqnarray}

Note that the operator $\overleftrightarrow{P}$ does not depend on $\vec{R}_\alpha$ because, owing to the periodic nature of the Bloch functions, the integral is the same for all $\vec{R}_\alpha$. Now, the envelope functions must be simultaneous solutions of this equation for all $\vec{R}_\alpha$. Thus, approximating $\vec{R}_\alpha$ as a continuous variable (i.e ignoring the discretization of the lattice of primitive unit cells) with the justifications that the envelope functions will vary slowly between unit cells, we can redefine the above equation into normal Schrodinger equation
\begin{equation}
\left(\overleftrightarrow{T}+\alpha\overleftrightarrow{P}\right)\cdot \vec{F}\left(\vec{r}\right)+\overleftrightarrow{\Delta V} \left(\vec{r}\right)\cdot \vec{F}\left(\vec{r}\right)=\Delta E \vec{F}\left(\vec{r}\right)
\end{equation}
Furthermore, we can simplify the definitions of the operators by 
\begin{eqnarray}
\overleftrightarrow{\Delta V}_{bb'}\left(\vec{r}\right)&=&\frac{1}{V_c}\int_{-\infty}^{\infty}f^*_{b'}\left(\vec{r'}\right)f_b\left(\vec{r'}\right)\Delta V\left(\vec{r'}+\vec{r}\right)d^3r'\\
\overleftrightarrow{P}_{bb'}&=&\frac{1}{V_c}\int_{-\infty}^{\infty}f^*_{b'}\left(\vec{r}\right)\vec{\nabla}f_b\left(\vec{r}\right)d^3r\cdot\vec{\nabla}
\end{eqnarray}
where $f_b\left(\vec{r}\right)$ is the Bloch function $u_b\left(\vec{r}\right)$ within the central unit cell and is enforced to be zero outside of that volume of the unit cell, thereby allowing the integral limits to be expanded to all space. 

\par The equations above do not yield the same energy spectrum as that of unperturbed diamond in the limit $\Delta V\rightarrow 0$. To rectify this, we need to include higher-order corrections that are not captured by the ansatz. In effective mass theory, this is achieved by redefining the kinetic energy operator with a mass that reproduces the dispersion relationship of each band. Typically, the coupling of bands by the $\overleftrightarrow{P}$ term is ignored for simplicity and to avoid ambiguity in the definition of the Bloch function basis for $\overleftrightarrow{\Delta V}$. We will implement both of these approximations to yield our effective mass equations
\begin{equation}
\left(\overleftrightarrow{T}_\text{eff}+\overleftrightarrow{\Delta V}\left(\vec{r}\right)\right)\cdot \vec{F}\left(\vec{r}\right)=\Delta E\vec{F}\left(\vec{r}\right)
\end{equation}
where $\overleftrightarrow{T}_\text{eff}$ is a diagonal tensor with elements that are the kinetic energy operator with the effective hole mass of the corresponding band.

\subsection{Simplest effective mass model of bound hole of NV$^-$}
\label{sec:simplestEM}
 In the simplest model, we will ignore any couplings between or differences in the potentials experienced by different bands, and approximate the Bloch function product as a single Gaussian in which
\begin{equation}
f^*_{b'}\left(\vec{r}\right)f_b\left(\vec{r}\right)=g\left(\lambda r^2\right)=\left(\lambda/\pi\right)^{3/2}V_c e^{-\lambda r^2}
\end{equation}
where $\lambda$ is the Gaussian orbital exponent to be determined. In this model, the elements of the potential tensor are
\begin{equation}
\overleftrightarrow{\Delta V}_{bb'}
\left(\vec{r}\right) =\delta_{bb'} \frac{1}{V_c}\int_{-\infty}^{\infty} g\left(\lambda r'^2\right) \Delta V\left(\vec{r'}+\vec{r}\right)d^3 r'
\end{equation}
Let us define the perturbation potential, in units of eV as
\begin{eqnarray}
\Delta V\left(\vec{r}\right)&=&-\kappa \sum_n \Delta q_n \frac{1}{\left|\vec{r}-\vec{R}_n\right|}+\kappa\sum_m Z_\text{C}\left(\frac{1}{\left|\vec{r}-\vec{R}_{\text{NV}^-,m}\right|}-\frac{1}{\left|\vec{r}-\vec{R}_{\text{d},m}\right|}\right) \nonumber\\ 
&&+\kappa\left(\frac{Z_\text{N}}{\left|\vec{r}-\vec{R}_{\text{NV}^-,\text{N}}\right|}-\frac{Z_\text{C}}{\left|\vec{r}-\vec{R}_{\text{d},\text{C}_\text{N}}\right|}\right)-\kappa\frac{Z_C}{\left|\vec{r}-\vec{R}_{\text{d},\text{C}_\text{V}}\right|}
\end{eqnarray}
where $\kappa=e/4\pi\epsilon$ and $\epsilon=5.7 \epsilon_0$ is the dielectric constant of diamond \cite{Madelung2004}. $\Delta q_n=\left[\rho_{\text{NV}^-}\left(\vec{R}_n\right)-\rho_\text{d}\left(\vec{R}_n\right)\right]V_\text{sampling}$ where $\rho_{\text{NV}^-}$ and $\rho_\text{d}$ are the valence electron densities for NV$^-$ and defect-free diamond, respectively. $V_\text{sampling}$ is the corresponding sampling volume which describes the sampling of the charge densities (i.e $V_\text{sampling}$= volume of simulation/ number of samples), $\vec{R}_n$ is the $n^{th}$ electron density sample position, $Z_\text{C}$ and $Z_\text{N}$ are the effective relative charges of the carbon and nitrogen nuclei, respectively. 

\par Now, the integral for one of the terms in the sum above can be evaluated analytically to be \cite{Szabo1996}
\begin{eqnarray}
\frac{\kappa}{V_c}\int_{-\infty}^{\infty}g\left(\lambda r^2\right) \frac{Z}{\left|\vec{r}-\vec{R}\right|}d^3r&=&\frac{\kappa}{V_c}V_c\left(\frac{\lambda}{\pi}\right)^{3/2}Z\frac{2\pi}{\lambda} F\left[\lambda\left|\vec{r}-\vec{R}\right|^2\right]\nonumber \\
&=&2\kappa Z\left(\frac{\lambda}{\pi}\right)^{1/2}F\left[\lambda\left|\vec{r}-\vec{R}\right|^2\right]\
\end{eqnarray}
where 

\begin{equation}
F\left[x\right]=\frac{1}{2}\sqrt{\frac{\pi}{x}}\text{erf}\left(\sqrt{x}\right)
\end{equation}
Thus, using this analytical result, the potential tensor can be written as 

\begin{eqnarray}
\overleftrightarrow{\Delta V}_{bb'}\left(\vec{r}\right)&=&\delta_{bb'}2\kappa\sqrt{\frac{\lambda}{\pi}}\Bigg\{
-\sum_n\Delta q_n F\left[\lambda\left|\vec{r}-\vec{R}_n\right|^2\right]+\sum_mZ_\text{C}\left(F\left[\lambda\left|\vec{r}-\vec{R}_{\text{NV}^-,m}\right|^2\right]-F\left[\lambda\left|\vec{r}-\vec{R}_{\text{d},m}\right|^2\right]\right)\nonumber\\
&&+\left(Z_\text{N}F\left[\lambda\left|\vec{r}-\vec{R}_{\text{NV}^-,\text{N}}\right|^2\right]-Z_\text{C}F\left[\lambda\left|\vec{r}-\vec{R}_{\text{d},\text{C}_\text{N}}\right|^2\right]\right)-Z_\text{C}F\left[\lambda\left|\vec{r}-\vec{R}_{\text{d},\text{C}_\text{V}}\right|^2\right]
\Bigg\}
\label{eq:potentialtensor}
\end{eqnarray}
As the potential is branch independent, the set of Schr\"odinger equations for each band is explicitly
\begin{equation}
\left(\vec{p}\cdot\overleftrightarrow{\frac{1}{2m_b}}\cdot\vec{p}+\Delta V_{b}\left(\vec{r}\right)\right)F_{b,i}\left(\vec{r}\right)=\Delta E_{b,i} F_{b,i}\left(\vec{r}\right)
\label{eq:finalSE}
\end{equation}
where $\Delta V_{b}\left(\vec{r}\right)=\overleftrightarrow{\Delta V}_{bb'}\left(\vec{r}\right)$ from above, $\overleftrightarrow{\frac{1}{m_b}}$ is the effective mass tensor of the $b^{th}$ band and the index $i$ denotes the $i^{th}$ level of each band. For simplicity, we assumed isotropic masses for the heavy holes and light holes, hence the final Schr\"odinger equation becomes
\begin{equation}
\left(\frac{\vec{p}^2}{2m_b}+\Delta V_{b}\left(\vec{r}\right)\right)F_{b,i}\left(\vec{r}\right)=\Delta E_{b,i} F_{b,i}\left(\vec{r}\right)
\label{eq:finalSEfinal}
\end{equation}

\section{Electron and nuclear charge potentials}
\par The calculation of the charge potentials requires us to estimate $\lambda$ in Eq.~\ref{eq:potentialtensor}. Since no definitive analytical form exists for the C orbitals of a multi-electron atom, we approximate the atomic orbitals of C using Slater type orbital (STO). As a result, we can approximate $\lambda$ by using a STO-1G model of carbon 2sp orbitals. The effective charge $Z_\text{eff}$ for 2s and 2p orbitals is 3.2166 and 3.1358, respectively \cite{Atkins2011}. We further approximated the effective charge by averaging them, giving us $Z_\text{eff}=3.1762$. From the Slater rules, the effective principal number $n_\text{eff}$ for $n=2$ is also 2. Hence, the final Slater orbital exponent for C can be approximated to be $\xi_C=3.1762/2=1.588$. 

\begin{figure}[h!]
\centering
    \begin{subfigure}[t]{0.9\textwidth}
    \includegraphics[width=\textwidth]{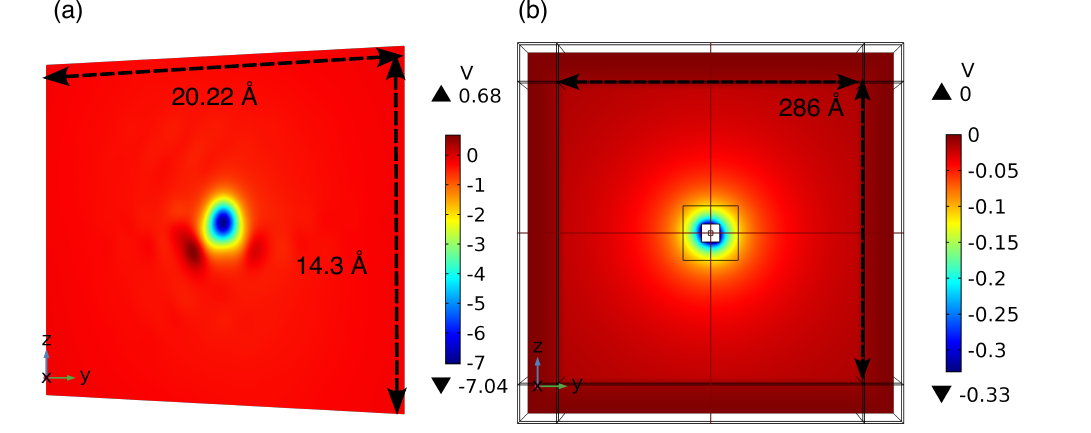}
    \phantomcaption
    \label{fig:nearfieldpotential}
    \end{subfigure}
    \begin{subfigure}[t]{0\textwidth}
    \includegraphics[width=\textwidth]{potential.pdf}
    \phantomcaption
    \label{fig:extrapolatedpotential}
    \end{subfigure}
\caption{(a) The near-field effective potential plotted over the (110) plane of a cubic unit cell. (b) The extrapolated near-field potential obtained by solving the Poisson's equation in COMSOL using infinite element domain. The blank space corresponds to the cubic unit cell.}
\label{fig:potential}
\end{figure}

\par The Gaussian orbital exponent scales as \cite{Szabo1996}
\begin{equation}
\lambda= \lambda\left(\xi=1.0\right)\times \xi_C^2
\end{equation}
Given $\xi=1.0$ for a hydrogen atom which corresponds to $\lambda=0.271$ for a single STO-1G description, the final Gaussian orbital exponent for C is then 0.683 (in atomic units).

\par Using the interpolated charge densities obtained from the $\textit{ab initio}$ calculations from Sec.~\ref{sec:chargedensityNV}, we construct $\Delta q_n$ with a grid dimension of $\left(x,y,z\right)\in \left(-7.15,7.15\right)~\text{\AA}$ with $V_\text{sampling}=14.3^3/192^3$. Similarly, we repeat the steps outlined above with the nuclear charges. We use $Z_\text{N}=5$ and $Z_\text{C}=4$ for the valence nuclear charges for the N and C atoms, respectively to construct the near-field effective potential, as shown in Fig.~\ref{fig:nearfieldpotential}.

\par After constructing the near-field effective potential, we extend this potential to encompass larger volumes to obtain the solutions of the envelope wavefunctions with a larger radius. This extrapolation of the near-field potential was performed by solving Poisson's equation using COMSOL Multiphysics\textsuperscript\textregistered version 5.3a. The resulting effective potential is shown in Fig.~\ref{fig:extrapolatedpotential}.

\section{Simulation of bound hole states}
\label{sec:simulateboundhole}
\par
\begin{figure}[h!]
\centering
    \begin{subfigure}[t]{0.9\textwidth}
    \includegraphics[width=\textwidth]{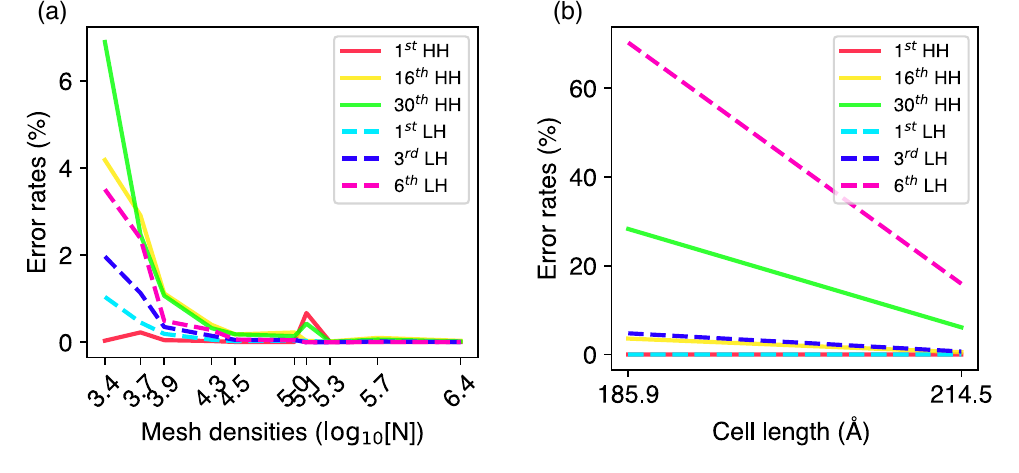}
    \phantomcaption
    \label{fig:meshconvergence}
    \end{subfigure}
    \begin{subfigure}[t]{0\textwidth}
    \includegraphics[width=\textwidth]{comsolconvergence.pdf}
    \phantomcaption
    \label{fig:cellconvergence}
    \end{subfigure}
\caption{(a) Mesh convergence studies of the HH and LH solutions, plotted with increasing mesh densities. The small bump at 5.1 is due to the further increment of the mesh densities close to the center of the unit cell where the effective potential varies the most [Fig.~\ref{fig:nearfieldpotential}]. (b) Cell convergence studies of the HH and LH solutions. The first errors rates are calculated based on the difference of the energy levels with $N = 2225$ mesh elements [Fig.~\ref{fig:meshconvergence}] and a cell length of 143~\AA [Fig.~\ref{fig:cellconvergence}], respectively.}
\label{fig:comsolconvergence}
\end{figure}

\begin{figure}[h!]
\centering
    \begin{subfigure}[t]{0.9\textwidth}
    \includegraphics[width=\textwidth]{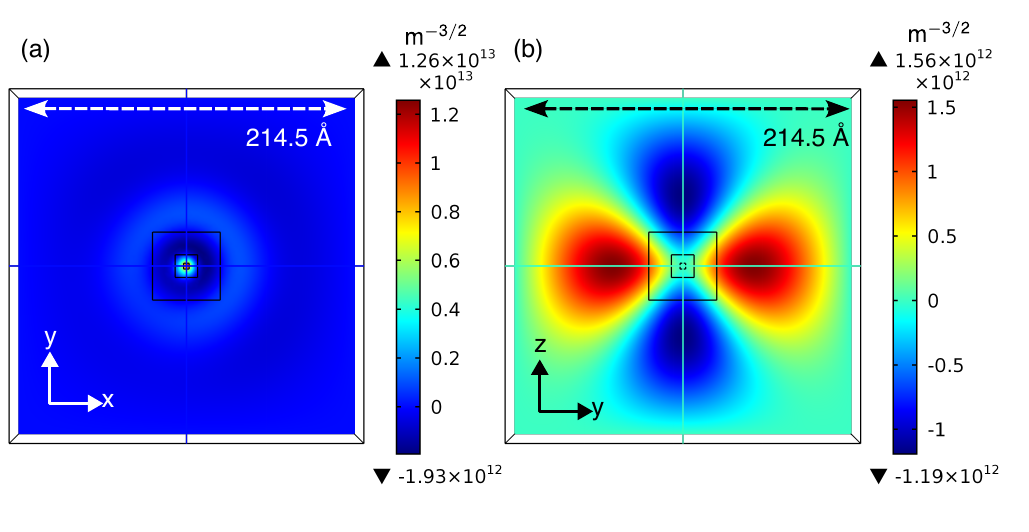}
    \phantomcaption
    \label{fig:HHlvl30}
    \end{subfigure}
    \begin{subfigure}[t]{0\textwidth}
    \includegraphics[width=\textwidth]{wfconfinement.pdf}
    \phantomcaption
    \label{fig:LHlvl10}
    \end{subfigure}
\caption{Artificial confinement of envelope wavefunctions. (a) Envelope wavefunctions for the 30$^{th}$ energy level of HH solutions (4s orbital) and (b) the 10$^{th}$ energy level of LH solutions (the fifth 3d orbital).}
\label{fig:wfconfinement}
\end{figure}

 We simulate the bound hole states in COMSOL Multiphysics\textsuperscript\textregistered~over a dimension of $214.5\times 214.5\times 214.5~\text{\AA}^3$ with a Dirichlet boundary condition and isotropic effective masses of $m_\text{HH}=1.08m_e$ and $m_\text{LH}=0.36m_e$ for the heavy and light hole mass, respectively where $m_e$ is the mass of the electrons \cite{Madelung2004}. To validate the accuracy of these calculations, we perform convergence studies by progressively increasing the mesh density and cell size. Our results show successful mesh convergence with deviations of less than 1\% for both HH and LH solutions and cell convergence with errors of less than 7\%(3\%) up to 30(5) levels for HH(LH) solutions, as shown in Fig.~\ref{fig:meshconvergence} and \ref{fig:cellconvergence}. Despite this, our findings indicate that the 3d orbital energy levels (6$^{th}$ to $10^{th}$ energy levels) of the LH solutions have cell convergence errors varying from 16\% to 25\% due to the strong artificial confinement as observed in Fig.~\ref{fig:LHlvl10}.

As illustrated in Fig.~\ref{fig:wfconfinement}, the envelope wavefunction of both of the bound states are artificially confined by the simulation domain. While it is possible to obtain higher excited bound hole states directly via COMSOL simulations, this approach is computationally inefficient as the wavefunctions of these bound states extend over tens of nanometers. To overcome this challenge, we spherically averaged the effective potential and solve for the radial Schr\"odinger equation to obtain the higher excited states. Tab~\ref{tab:energylevelsCOMSOL} shows the comparison between the spherically averaged solutions and the COMSOL solutions. The COMSOL solutions exhibit additional splittings due to the accurate representation of the defect's true $C_{3v}$ symmetry, which is absent in the spherically averaged solutions. By ignoring the splittings and considering the mean values of the energy levels, we find that HH and LH solutions of these two methods differ by less than 3\%, up to 30 levels for HH and 5 levels for LH, respectively. While the HH solutions are in good agreement, the LH solutions for 3d orbitals and beyond are subject to inaccuracies caused by the strong artificial confinement of the envelope wavefunctions [Fig.~\ref{fig:LHlvl10}], resulting in inaccurate predictions of the energy levels. A more accurate representation of the energy levels can only be achieved by using a larger simulation domain, thus avoiding the artificial confinement of the envelope wavefunctions.
\begin{table}
\caption{\label{tab:energylevelsCOMSOL} Energy levels of the bound hole states. The underlined values represent the average of the energy levels within that manifold. The second and third column correspond to the HH energy levels obtained from COMSOL using the full effective potential and the one-dimensional spherically averaged potential, respectively. The fourth column shows the difference between the averaged energy values obtained from COMSOL and the spherically averaged potential solutions. The fifth to the seventh column corresponds to the LH solutions.}
\begin{ruledtabular}
\begin{tabular}{ccccccc}
Energy levels &  $E_{\text{HH},\text{COMSOL}}$ (eV) &  $E_{\text{HH},\text{sph}}$ (eV) & $\Delta E_{\text{HH}}$ (\%) &$E_{\text{LH},\text{COMSOL}}$ (eV) &$E_{\text{LH},\text{sph}}$ (eV) &  $\Delta E_{\text{LH}}$ (\%)\\ \hline
1s &-0.24346  & -0.23928 & 1.717&-0.11891 & -0.11830 & 0.515\\ 
\hline 
\multirow{3}{*}{2p} 
& -0.10382& -0.10291 & 0.874 &-0.03723  &-0.03701   &0.583\\
\cline{2-2}
\cline{5-5}
& -0.10620& & & -0.03754 &  &  \\
& -0.10619& & &-0.03754 & & \\
& -0.09907& & &-0.03661 & & \\

\hline 
2s & -0.08440& -0.08325 & 1.363 &-0.03285 & -0.03307 & -0.676 \\
\hline 
\multirow{5}{*}{3d} 
& -0.04992 & -0.04948 & 0.878&-0.01361 & -0.01643 &-20.697  \\
\cline{2-2}
\cline{5-5}
& -0.05078& & &-0.01422 &  &  \\
& -0.05078& & &-0.01422 & & \\
& -0.04940& & &-0.01414 & & \\
& -0.04940& & &-0.01274 & & \\
& -0.04924& & &-0.01274 & & \\

\hline 
\multirow{3}{*}{3p} 
& -0.04706 & -0.04667 & 0.826 && & \\
\cline{2-2}
& -0.04763 & &  &&  &  \\
& -0.04763 & &  && & \\
& -0.04593 & &  && & \\
\hline 

3s &-0.04130 & -0.04084 & 1.133&  & &  \\ 
\hline 

\multirow{7}{*}{4f}
& -0.02804 & -0.02783 & 0.746 &  &  & \\
\cline{2-2}
& -0.02822& & & &  &  \\
& -0.02822& & & & & \\
& -0.02802& &  && & \\
& -0.02802& &  && & \\
& -0.02794& & & & & \\
& -0.02794& & & & & \\
& -0.02792& & & & & \\
\hline 

\multirow{5}{*}{4d}
& -0.02788&  -0.02776& 0.432& &  & \\
\cline{2-2}
 & -0.02826& & & &  &  \\
& -0.02826& &  && & \\
& -0.02766& & & & & \\
& -0.02760& & & & & \\
& -0.02760& & & & & \\
\hline 

\multirow{3}{*}{4p}
& -0.02645& -0.02655 & -0.356&& & \\
\cline{2-2}
& -0.02669& &  &&  &  \\
& -0.02669& &  && & \\
& -0.02597& & & & & \\

\hline 
4s & -0.02355 & -0.02405 & -2.132 & &
\end{tabular}
\end{ruledtabular}
\end{table}

\newpage

\section{Deformation potential model hole-phonon scattering}
\label{sec:scatteringrate}
\subsection{Hole-acoustic phonon scattering}
\label{sec:acousticscattering}
Hole-phonon scattering is the process in which the energy of a hole is altered slightly by a phonon interaction. This scattering rate can be modelled using Fermi's golden rule \cite{Ridley2013}. The acoustical deformation potential interaction is given by
\begin{equation}
\hat{H}_\text{ac}=\Theta \vec{\nabla}\cdot \vec{U}
\end{equation}
where $\Theta = 5.5$~eV is the hole acoustic-phonon deformation potential \cite{Jacoboni1983} and $\vec{U}$ is the phonon displacement field. Introducing the phonon modes, the potential becomes
\begin{equation}
\hat{H}_\text{ac}=\Theta \sum_{\beta,\vec{k}} \vec{\nabla}\cdot \vec{U}_{\beta,\vec{k}} A_{\beta,\vec{k}}
\end{equation}
where $\vec{U}_{\beta,\vec{k}}$ is the normalized displacement field associated with the mode of branch $\beta$ and wavevector $\vec{k}$, and $A_{\beta,\vec{k}}$ is the amplitude of the mode. In this case, the amplitudes can be written in terms of the phonon creation (annihilation) operator $a^\dagger_{\beta,\vec{k}}$ $\left(a_{\beta,\vec{k}}\right)$
\begin{equation}
A_{\beta,\vec{k}}=\sqrt{\frac{\hbar}{2\rho V\omega_{\beta,\vec{k}}}}\left(a_{\beta,\vec{k}}+a^\dagger_{\beta,\vec{k}}\right)
\label{eq:phononamplitudes}
\end{equation}
where $\rho = 3.515 \text{g/cm}^3$ is the density of the diamond \cite{Madelung2004} and $\omega_{\beta,\vec{k}}$ is the mode frequency. For bulk diamond, the only modes that contribute to the deformation potential are the longitudinal modes because transverse modes satisfy $\vec{\nabla}\cdot \vec{U}=0$. Thus, we can reduce the sum over mode branches in the above to just the longitudinal modes. As a consequence, we can derive the longitudinal modes from a scalar field $\psi_{\vec{k}}$ where
\begin{equation}
\vec{U}_{\vec{k}}=\vec{\nabla}\psi_{\vec{k}}
\end{equation}
with the dispersion relation of 
\begin{equation}
\omega_k=c_lk
\end{equation}
where $c_l \approx 18330~\text{ms}^{-1}$ is the longitudinal speed of sound in diamond \cite{Madelung2004,McSkimin1972}. This scalar potential satisfies the wave equation 
\begin{equation}
-\nabla^2 \psi_{\vec{k}}=\frac{\omega^2}{c_l^2} \psi_{\vec{k}}
\end{equation}
The solution to $\psi_{\vec{k}}$ can be determined via the normalization condition in which
\begin{equation}
\int_V \vec{U}^*_{\beta,\vec{k}}\cdot\vec{U}_{\beta',\vec{k}'} d^3r=V\delta_{\beta,\beta'}\delta_{\vec{k},\vec{k}'}
\label{eq:phononnormalization}
\end{equation}
where $V$ is the volume of the diamond. In this case, we have
\begin{equation}
\psi_{\vec{k}}=\frac{c_l}{\omega_k}e^{\mathfrak{i}\vec{k}\cdot\vec{r}}
\end{equation}
and 
\begin{equation}
\vec{\nabla}\cdot\vec{U}_{\vec{k}}=\nabla^2\psi_{\vec{k}}=-\frac{\omega_k}{c_l}e^{\mathfrak{i}\vec{k}\cdot\vec{r}}
\end{equation}
Finally, the acoustic deformation potential in bulk diamond is
\begin{equation}
\hat{H}_\text{ac}=\frac{\Xi}{\sqrt{V}}\sum_{\vec{k}}\sqrt{\omega_k} \left(a_{\vec{k}}+a^\dagger_{\vec{k}}\right)e^{\mathfrak{i}\vec{k}\cdot\vec{r}}
\end{equation}
where $\Xi=\Theta\sqrt{\hbar/2\rho c_l^2}$.
\par The holes are transitioning between the different bound states confined by a one-dimensional potential as obtained from our previous calculations. Now, let us consider the transition rate for the hole to be excited out of the $i^{th}$ level of the $b^{th}$ band, $F_{b,i}$ to a higher energy state through the absorption of one phonon. Letting the energy difference of the different transitions to be $\Delta E_{b,i}^{b',i'}=E_{b',i'}-E_{b,i}$ and using Fermi's golden rule \cite{Ridley2013}, we have
\begin{equation}
\Gamma_\text{ac}\ _{{b,i}}^{b',i'} =\frac{2\pi}{\hbar} \sum_{b',i',\vec{k}}\left|\left<n_{\vec{k}}-1\left|\left<F_{b',i'}^*\left|\hat{H}_\text{ac}\right|F_{b,i}\right>\right|n_{\vec{k}}\right>\right|^2 \delta\left(\Delta E_{b,i}^{b',i'}-\hbar\omega_k\right)
\end{equation}
where we are summing over all the possible levels $i$ and the different hole bands $b$. Making the substitution for $\hat{H}_\text{ac}$ and perform inner products, we arrive at 
\begin{equation}
 \Gamma_\text{ac}\ _{{b,i}}^{b',i'}  =\frac{\Xi^2}{V}\frac{2\pi}{\hbar} \sum_{b',i',\vec{k}}\omega_k\ n_{\vec{k}}\left|\left<F_{b',i'}^*\left|e^{\mathfrak{i}\vec{k}\cdot\vec{r}}\right|F_{b,i}\right>\right|^2 \delta\left(\Delta E_{b,i}^{b',i'}-\hbar\omega_k\right)
 \end{equation}
Average over the thermal distribution of initial phonon states, we have
\begin{eqnarray} 
\Gamma_\text{ac}\ _{{b,i}}^{b',i'}  &=&\frac{\Xi^2}{V}\frac{2\pi}{\hbar} \sum_{b',i',\vec{k}}\omega_k\  n_B\left(\omega_k,T\right) \left|\left<F_{b',i}^*\left|e^{\mathfrak{i}\vec{k}\cdot\vec{r}}\right|F_{b,i}\right>\right|^2 \delta\left(\Delta E_{b,i}^{b',i'}-\hbar\omega_k\right)\nonumber \\ 
 &=&\frac{\Xi^2}{V}\frac{2\pi}{\hbar} \sum_{b',i',\vec{k}}\omega_k\  n_B\left(\omega_k,T\right) \left|M_{b,i}^{b',i'}\right|^2 \delta\left(\Delta E_{b,i}^{b',i'}-\hbar\omega_k\right)
\end{eqnarray}
where $n_B\left(\omega_k,T\right)$ is the Bose-Einstein distribution and we have defined the matrix element $M_{b,i}^{b',i}$ as
\begin{equation}
M_{b,i}^{b',i'}\left(\vec{k}\right)= \int_{0}^{\infty}\int_{0}^{\infty}\int_{0}^{\infty} F_{b',i'}^* F_{b,i} e^{\mathfrak{i}\vec{k}\cdot\vec{r}} dx dy dz
\end{equation}
Furthermore, we can convert the discrete sum over the wavevector $\vec{k}$ into a continuous form by introducing a normalization constant of $V/\left(2\pi\right)^3$, giving us 
\begin{eqnarray}
\Gamma_\text{ac}\ _{{b,i}}^{b',i'} &=&\frac{\Xi^2}{V}\frac{2\pi}{\hbar}\frac{V}{\left(2\pi\right)^3}\sum_{b',i'}\int \omega_k\ n_B\left(\omega_k,T\right) \left|M_{b,i}^{b',i'}\right|^2\delta\left(\Delta E_{b,i}^{b',i'}-\hbar\omega_k\right) d^3k\nonumber \\
&=&\frac{\Theta^2}{2\left(2\pi\right)^2\rho c_l^2}\sum_{b',i'}\int \omega_k\ n_B\left(\omega_k,T\right) \left|M_{b,i}^{b',i'}\right|^2\delta\left(\Delta E_{b,i}^{b',i'}-\hbar\omega_k\right) d^3k
\label{eq:ackspace}
\end{eqnarray}
Eq.~\ref{eq:ackspace} can be simplified by converting the integral over $k$-space into spherical coordinates. Using this simplification, we reapply the dispersion relation to obtain
\begin{equation}
\Gamma_\text{ac}\ _{{b,i}}^{b',i'} =\frac{\Theta^2}{2\left(2\pi\right)^2\rho c_l^5}\sum_{b',i'}\int \omega_k^3\ n_B\left(\omega_k,T\right) \left|M_{b,i}^{b',i'}\right|^2\delta\left(\Delta E_{b,i}^{b',i'}-\hbar\omega_k\right)\sin\theta_k d\omega_k d\theta_k d\phi_k
\label{eq:acbeforedirac}
\end{equation}
Employing the properties of the Dirac delta distribution, the general expression for a hole-acoustic phonon scattering rate is then
\begin{equation}\Gamma_\text{ac}\ _{{b,i}}^{b',i'} =\frac{\Theta^2}{2\left(2\pi\right)^2\rho\hbar c_l^5}\sum_{b',i'}\int \Delta \omega_{b',i'}^3\ n_B\left( \Delta \omega_{b',i'},T+\frac{1}{2}\mp \frac{1}{2}\right) \left|M_{b,i}^{b',i'}\right|^2\sin\theta_k  d\theta_k d\phi_k\\
\end{equation}
where the $+$ and $-$ sign denotes the emission and absorption processes, respectively. We have also expressed the matrix element in terms of spherical coordinates
\begin{eqnarray}
M_{b,i}^{b',i}\left(\Delta\omega_{b',i'},\theta_k,\phi_k\right)&=&\int_{0}^{R}\int_{0}^{\pi}\int_{0}^{2\pi} F_{b,i}^* F_{b,i}r^2 \sin \theta \nonumber \\ 
&\times & e^{ir\Delta\omega_{b',i'}/c_l\left(\sin\theta\sin\theta_k\cos\phi\cos\phi_k+\sin\theta\sin\theta_k\sin\phi\sin\phi_k+\cos\theta\cos\theta_k\right)}dr d\theta d\phi
\label{eq:acfinalmatrixelement}
\end{eqnarray}
In the above calculations, the angular integrals must be evaluated numerically and we need to ensure the wavefunctions are normalized where the envelope wavefunctions are expressed in terms of both radial and spherical harmonics components
\begin{eqnarray}
F_{b,i}&=&R_{b,n_i,l_i}Y_{l_i,m_{l_i}}\nonumber \\
1&=&\int \left|R_{b,n_i,l_i}\right|^2\left|Y_{l_i,m_{l_i}}\right|^2 r^2\sin\theta dr d\theta d\phi
\end{eqnarray}

\subsection{Hole-optical phonon scattering}
\label{sec:opticalscattering}
\begin{figure}[h!]
\centering
\includegraphics[width=0.45\textwidth]{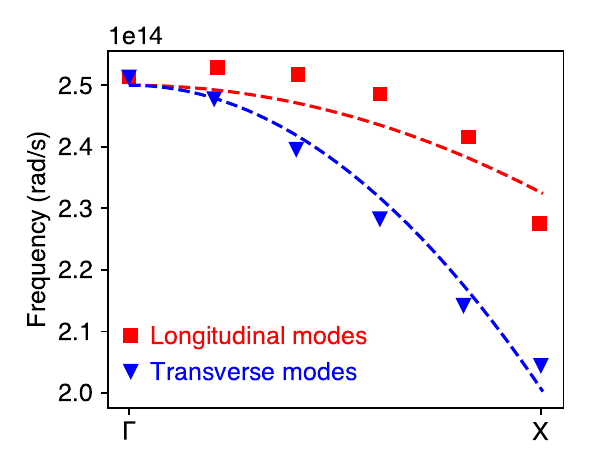}
\caption{Optical phonon dispersion curves of diamond. The solid markers represent the experimental data obtained from Ref.~\cite{Warren1967,Weber1977} while the dashed lines are fitted using Eq.~\ref{eq:opdispersionrelation}. The curvature for longitudinal and transverse optical phonon modes is $\eta_{\parallel}=5.619\times 10^{-8}~\text{rad m}^2/\text{s}$ and $\eta_\perp=1.588\times 10^{-7}~\text{rad m}^2/\text{s}$, respectively.}
\label{fig:phononfits}
\end{figure}

For hole-optical phonon scattering, the optical deformation potential interaction is given by \cite{Ridley2013}
\begin{equation}
\hat{H}_\text{op}=\frac{1}{2}\vec{D}\cdot \vec{U}
\end{equation}
where $\vec{D}$ is the optical deformation potential vector. Introducing the phonon modes, we have
\begin{equation}
\vec{u}=\sum_{\beta,\vec{k}}\vec{U}_{\beta,\vec{k}} A_{\beta,\vec{k}}
\end{equation}
where $\vec{U}$ is the unit mode displacement normalized using Eq.~\ref{eq:phononnormalization}. Similarly, we then quantize the displacement amplitudes using Eq.~\ref{eq:phononamplitudes}. Combining all together, we have
\begin{eqnarray}
\hat{H}_\text{op}&=& \frac{1}{2}\sum_{\beta,\vec{k}}\vec{D}\cdot\vec{U}_{\beta,\vec{k}}A_{\beta,\vec{k}}\nonumber \\
&=&\frac{1}{2}\sum_{\beta,\vec{k}}\left(D_0 \hat{D}\right)\cdot\left(\left(\hat{k}\times \hat{n}\right)e^{i\vec{k}\cdot\vec{r}}\right)A_{\beta,\vec{k}}\nonumber \\
&=&\frac{1}{2}\sum_{\beta,\vec{k}}D_0\cos\theta_{\vec{D}\cdot\vec{U}_{\beta,\vec{k}}}e^{i\vec{k}\cdot\vec{r}}A_{\beta,\vec{k}}
\end{eqnarray}
where $\hat{D}\cdot \left(\hat{k}\times\hat{n}\right)=\cos\theta_{\vec{D}\cdot\vec{U}_{\beta,\vec{k}}}$ and $\theta_{\vec{D}\cdot\vec{U}_{\beta,\vec{k}}}$ is the angle between the optical deformation potential and various phonon modes. The dispersion relation is given by
\begin{equation}
\omega_{\beta,k}=\omega_0 -\eta_\beta k^2
\label{eq:opdispersionrelation}
\end{equation}
where $\omega_0=2.5\times 10^{12}~\text{rad/s}$ is the frequency which corresponds to the highest phonon energy in diamond \cite{Madelung2004}, $\eta_\beta$ is the curvature of the optical phonon dispersion curves of various modes [Fig.~\ref{fig:phononfits}]. Finally, the optical deformation potential in bulk diamond is
\begin{equation}
\hat{H}_\text{op}=\frac{\Xi}{2\sqrt{V}}\sum_{\vec{k}} \frac{1}{\sqrt{\omega_{\beta,k}}} \cos\theta_{\vec{D}\cdot\vec{U}_{\beta,\vec{k}}} \left(a_{\beta,\vec{k}}+a^\dagger_{\beta,\vec{k}}\right)e^{\mathfrak{i}\vec{k}\cdot\vec{r}} 
\end{equation}
where $\Xi=D_0 \sqrt{\hbar/2\rho}$. Let us consider the transition rate for the hole to be de-excited out of a higher energy level to the lower energy state through the emission of one phonon.  We have
\begin{equation}
 \Gamma_\text{op}\ _{{b',i'}}^{b,i} =\frac{2\pi}{\hbar} \sum_{\beta,b',i',\vec{k}}\left|\left<n_{\beta,\vec{k}}+1\left|\left<F_{b',i'}^*\left|\hat{H}_\text{op}\right|F_{b,i}\right>\right|n_{\beta,\vec{k}}\right>\right|^2 \delta\left(\Delta E_{b',i'}^{b,i}+\hbar\omega_{\beta,k}\right)
 \end{equation}
Making the substitution for $\hat{H}_\text{op}$, perform inner products and average over the thermal distribution of the initial phonon states, we have
\begin{equation}
\Gamma_\text{op}\ _{{b',i'}}^{b,i} =\frac{\Xi^2}{4V}\frac{2\pi}{\hbar} \sum_{\beta,b',i',\vec{k}}\frac{1}{\omega_{\beta,k}}\ \Bigg(n_{B}\left(\omega_{\beta,k},T\right)+1\Bigg)\left|\cos\theta_{\vec{D}\cdot\vec{U}_{\beta,\vec{k}}}\right|^2\left|\left<F_{b',i'}^*\left|e^{\mathfrak{i}\vec{k}\cdot\vec{r}}\right|F_{b,i}\right>\right|^2 \delta\left(\Delta E_{b',i'}^{b,i}+\hbar\omega_{\beta,k}\right)
\end{equation}
Following the similar procedures in the derivation of hole-acoustic phonon scattering, i.e Eq.~\ref{eq:ackspace}-\ref{eq:acbeforedirac}, we have
\begin{eqnarray}
\Gamma_\text{op}\ _{{b',i'}}^{b,i}&=&\frac{D_0^2}{8\left(2\pi\right)^2\rho\hbar} \sum_{\beta,b',i'}\int\frac{1}{\omega_{\beta,k}}\ \Bigg(n_{B}\left(\omega_{\beta,k},T\right)+1\Bigg)\left|\cos\theta_{\vec{D}\cdot\vec{u}_{\beta,\vec{k}}}\right|^2\nonumber \\ 
&&\times \left|M_{b',i'}^{b,i}\right|^2 \delta\left(\omega_{\beta,k}-\Delta\omega_{b',i'}\right)k^2_\beta \sin\theta_{\beta,k}dk_{\beta}d\theta_{\beta,k}d\phi_{\beta,k}
\end{eqnarray}
We use chain rules to to evaluate the integral over $dk_{\beta}$ where
\begin{eqnarray}
dk_\beta&=&\frac{dk_\beta}{d\omega_{\beta,k}}d\omega_{\beta,k}\nonumber \\
&=&d\omega_{\beta,k}\frac{1}{2\sqrt{\eta_{\beta}\left(\omega_0-\omega_{\beta,k}\right)}}
\end{eqnarray}
in which we have used the negative branch of $k_\beta$. Thus, we have

\begin{eqnarray}
\Gamma_\text{op}\ _{{b',i'}}^{b,i}&=&\frac{D_0^2}{8\left(2\pi\right)^2\rho\hbar} \sum_{\beta,b',i'}\int\ \Bigg(n_{B}\left(\omega_{\beta,k},T\right)+1\Bigg)\sin\theta_{\beta,k}\left|\cos\theta_{\vec{D}\cdot\vec{u}_{\beta,\vec{k}}}\right|^2\nonumber \\ 
&&\times \left|M_{b',i'}^{b,i}\right|^2 \delta\left(\omega_{\beta,k}-\Delta\omega_{b',i'}\right)\frac{\sqrt{\omega_0-\omega_{\beta,k}}}{2\eta_\beta^{3/2}\omega_{\beta,k}}d\omega_{\beta,k}d\theta_{\beta,k}d\phi_{\beta,k}\nonumber \\
&=&\frac{D_0^2}{8\left(2\pi\right)^2\rho\hbar} \sum_{\beta,b',i'}\int\ \Bigg(n_{B}\left(\omega_{\beta,k},T\right)+1\Bigg)\sin\theta_{\beta,k}\left|\cos\theta_{\beta,k}\right|^2\nonumber \\ 
&&\times \left|M_{b',i'}^{b,i}\right|^2 \delta\left(\omega_{\beta,k}-\Delta\omega_{b',i'}\right)\frac{\sqrt{\omega_0-\omega_{\beta,k}}}{2\eta_\beta^{3/2}\omega_{\beta,k}}d\omega_{\beta,k}d\theta_{\beta,k}d\phi_{\beta,k}
\end{eqnarray}
where in the last step we have simplified our cosine expression by defining our $z$ coordinate to be parallel to $\vec{D}$. Furthermore, we can relate the scalar magnitude $D_0$ to $d_0$ as follows \cite{Ridley2013}
\begin{equation}
D_0^2=\frac{3}{2}\frac{d_0^2}{a_0^2}
\end{equation}
where $d_0=61.2~\text{eV}$ is the deformation potential constant for hole-optical phonon interaction \cite{Jacoboni1983} and $a_0=3.567~\text{\AA}$ is the lattice constant of diamond \cite{Madelung2004}.  Evaluating the integral over $d\omega_{\beta,k}$ and taking into account the absorption/emission process, the general expression for a hole-optical phonon scattering rate is
\begin{equation}
\Gamma_\text{op}\ _{{b',i'}}^{b,i}=\frac{3d_0^2}{16\left(2\pi\right)^2 a_0^2\rho\hbar} \sum_{\beta,b',i'}\int\ \Bigg(n_{B}\left(\Delta\omega_{b',i'},T\right)+\frac{1}{2}\mp \frac{1}{2}\Bigg)\sin\theta_{\beta,k}\left|\cos\theta_{\beta,k}\right|^2\left|M_{b',i'}^{b,i}\right|^2 \frac{\sqrt{\omega_0-\Delta\omega_{b',i'}}}{2\eta_\beta^{3/2}\Delta\omega_{b',i'}}d\theta_{\beta,k}d\phi_{\beta,k}
\end{equation}
The final matrix element $M$ is 
\begin{eqnarray}
M_{b',i'}^{b,i}\left(\Delta\omega_{b',i'},\theta_{\beta,k},\phi_{\beta,k}\right)&=& \int_{0}^{\infty}\int_{0}^{\pi}\int_{0}^{2\pi} F_{b',i'}^*F_{b,i}r^2 \sin \theta\nonumber  \\
&&\exp{\Bigg[\mathfrak{i}r\sqrt{\frac{\omega_0-\Delta\omega_{b',i'}}{\eta_\beta}}\left(\sin\theta\sin\theta_{\beta,k}\cos\phi\cos\phi_{\beta,k}+\sin\theta\sin\theta_{\beta,k}\sin\phi\sin\phi_{\beta,k}+\cos\theta\cos\theta_{\beta,k}\right)\Bigg]}\nonumber \\
&&dr d\theta d\phi
\end{eqnarray}

\subsection{Scattering rate calculations}
\begin{figure}[h!]
\centering
\includegraphics[width=0.45\textwidth]{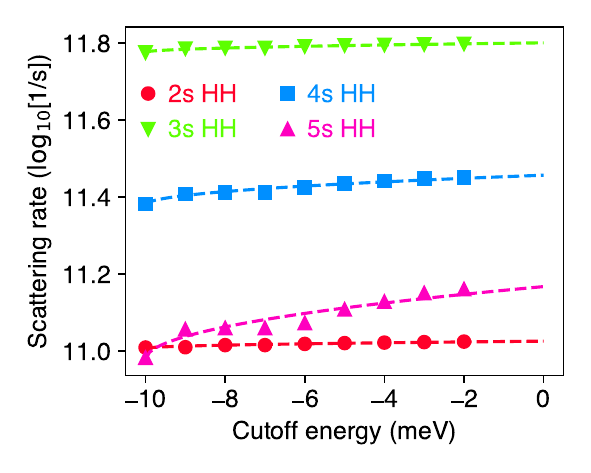}
\caption{Total absorption rates from 2s, 3s, 4s and 5s HH states at $T=300~$K. The solid markers represent the calculate values whereas the dashed lines are fitted using $a\times \sqrt{\left(E-10\right)}+b$ where we have used the $-10~$meV cutoff energy as a reference level. The energy level of 0~meV corresponds to the energy level of the VBM, which is at the commencement of a continuum levels of valence band states occupied by a hole. The fitting parameters $(a,b)$ for 2s to 5s HH absorption rates are given by $(0.0059,11.0076)$, $(0.0077,11.7762)$, $(0.0243,11.3801)$ and $(0.0599,10.9786)$, respectively.}
\label{fig:absorptionfits}
\end{figure}
\par Having derived the expressions for the hole-acoustic and hole-optical phonon scattering rates, we now calculate these scattering rates for all possible direct transitions from and to the GS, first excited state (ES), and $n$s HH states $\left(n=2-5\right)$. The longitudinal acoustic phonons exhibit non-linear dispersion relationship for energies above 70~meV \cite{SchwoererBoehning1998}. Consequently, based on the energy levels calculated in Sec~\ref{sec:simulateboundhole} and to a good approximation, we assume a range of $[0,120]$~meV for acoustic phonon modes energies to maximize the contribution to $\Gamma_\text{ac}$ and a range of $(120,165]$~meV for optical phonon modes \cite{Warren1967,Weber1977}.

\par The wavefunctions of higher excited states are more diffuse, leading to smaller overlaps between the envelope wavefunctions and therefore, decreasing scattering rates. Additionally, the phonon occupation factor causes the scattering rate to decrease as phonon energy increases. However, the increased density of states leads to more possible direct transitions. To calculate the total absorption rate from a particular energy level, we fit a square root function and converge the absorption rates by gradually decreasing the cutoff energies below the VBM. Fig.~\ref{fig:absorptionfits} shows the absorption rates from the calculated energy levels and Tab.~\ref{tab:scatteringrate} provides the total scattering rates. In general, there is a decreasing trend in both absorption and emission rates for higher excited states, as lower excited states that would contribute to larger matrix elements are omitted. Nevertheless, we note that the absorption rate from the 2s HH level and the emission rate from the 1s LH are lower, which may be attributed to the resonance of the wavevector operator $\left(e^{i\vec{k}\cdot\vec{r}}\right)$ with the envelope wavefunctions. Moreover, the combined effect of acoustic and optical phonon scattering results in a significantly higher total emission rate from the 2s HH state. 

\begin{table}[h]
\caption{Scattering rates from various energy levels at $T=300~$K and 5~K (in bracket). The absorption and emission rates denote the total rates out from the specific energy level.} 
\label{tab:scatteringrate}
\begin{ruledtabular}
\begin{tabular}{cccc}
 Energy levels & Total absorption rate (THz) & Total emission rate (THz) & Total scattering rate (THz)\\ \hline
 1s HH (GS) & 0.010 (0) & - & 0.010 (0) \\
 1s LH ($1^{st}$ ES)& 2.026 (0) & 0.015 (0.015) & 2.041 (0.015) \\ 
 2s HH & 0.106 (0) & 6.302 (5.482) & 6.408 (5.482) \\
 3s HH & 0.611 $(\approx 2\times 10^{-6})$ & 0.905 (0.449) & 1.516 (0.449) \\ 
 4s HH & 0.282 $(\approx 2\times 10^{-8})$ & 0.536 (0.315) & 0.818 (0.315) \\ 
 5s HH & 0.145 $(\approx 4\times 10^{-5})$ & 0.250 (0.149) & 0.395 (0.149) \\ 
\end{tabular}
\end{ruledtabular}
\end{table}
\begin{figure}[h!]
\centering 
\includegraphics[width=0.5\textwidth]{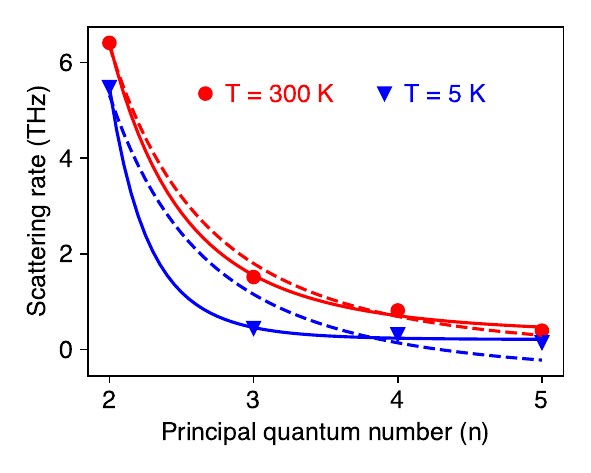}
\caption{Total scattering rates of 2s to 5s HH energy levels. The dashed lines are the fitted using $a+b/n^3$ whereas the solid lines are fitted with $c+d/n^e$ where $a,b,c,d$ and $e$ are the fitting parameters. Red and blue color correspond to $T=300$~K and $T=5$~K, respectively. The obtained fits for $T=300$~K are : -0.115, 51.702, 0.298, 91.263 and 3.901, which corresponds to $a,b,c,d$ and $e$, respectively. Similarly, the fits for $T=5$~K are : -0.598, 47.330, 0.209, 952.591 and 7.497.}
\label{fig:linewidthdependence}
\end{figure}

\par Fig.~\ref{fig:linewidthdependence} illustrate how the linewidths (or scattering rates) depend on the principal quantum number $n$. In the case of atoms, the primary mechanism governing linewidths is the radiative relaxation process. However, in materials, relaxation can be influenced by multiple factors, including both the radiative relaxation process (such as radiative recombination) and electron-phonon scattering \cite{Kazimierczuk2014,Heckoetter2017,Assmann2020}. When considering only radiative recombination, one expects a scaling law that goes as $n^{-3}$ \cite{Assmann2020}. In contrast, different types of electron-phonon scattering processes can result in distinct scaling laws for linewidths at different temperatures. For instance, in cuprous oxide, Rydberg excitons are known to interact with both longitudinal acoustic phonons via deformation potential coupling and longitudinal optical phonons via the Fr\"{o}hlich interaction \cite{Schweiner2016}. At low temperatures, it has been found that the contributions from the acoustic phonons are negligible \cite{Stolz2018}. Furthermore, recent study have also revealed that the linewidths of excitons in cuprous oxide are predominantly influenced by the longitudinal optical phonons via deformation potential coupling at low temperatures \cite{Stolz2018}, in contrast to the findings of Kitamura et al. \cite{Kitamura2017}. Additionally, the yellow $n$p Rydberg excitons are able to relax to the yellow $1$s states via optical phonon emission for $n$ as high as 30 \cite{Versteegh2021}. The linewidth broadening due to optical phonon emission in cuprous oxide exhibits a scaling law of $n^{-3}$ at low temperatures \cite{Stolz2018}.

\par In contrast to a bound exciton, which comprises an electron-hole pair orbiting each other and possesses momentum, in this work, we are considering a stationary bound hole orbiting the NV$^-$ defect and do not posses the aforementioned momentum. The Coulombic interaction between the negatively charged defect and the positively charged hole results in the hole being attracted towards and orbiting the defect, thereby creating localized electronic states. As a result, the interactions of a hole orbiting a negatively charged defect and phonons will exhibit fundamental differences compared to the interactions involving a Rydberg exciton. 

\par The sum of the radiative and non-radiative relaxation rates constitutes the total relaxation rate of the bound hole states. The radiative relaxation processes involve transitions to the internal levels of NV$^0$. Due to the more delocalized nature of the bound hole states, the dipole moment of the bound hole states will be smaller compared to the internal transitions of NV$^0$. As a consequence, the radiative relaxation rates of the bound hole states will be smaller. We assume that any non-radiative relaxation from the bound hole state to the excited state of NV$^0$ occurs at an equally slow rate. The NV defect has an internal radiative relaxation rates on the order of 10~ns \cite{Doherty2013}, which will be slower for the loosely bound states as previously discussed. As seen in Tab.~\ref{tab:scatteringrate}, the phonon scattering rates are considerably faster than the internal radiative relaxation rate of the NV center. Consequently, the primary mechanism responsible for broadening the linewidths of a bound hole captured by the NV$^-$ center is the scattering of phonons, which is distinct from Rydberg excitons in cuprous oxide where both radiative relaxation and phonon scattering mechanisms are equally important \cite{Kazimierczuk2014,Stolz2018,Assmann2020}. 

 \par It was identified that longitudinal optical phonons do not contribute to the scattering rates for $n>2$ HH states due to the range of the available optical phonon energies. For $n\geq 3$ HH states, only longitudinal acoustic phonons contribute to the first order scattering rate calculations and there are no direct transitions to the ground state. This contrasts with the Rydberg excitons in cuprous oxide, where the contribution of acoustic phonons via deformation potential coupling is insignificant. Since optical phonons contribute significantly to the linewidths, the exclusion of this mode in the scattering rates results in smaller scattering rates, as observed in the non-cubic inverse power laws as depicted in Fig.~\ref{fig:linewidthdependence}.

\par Thus, it is anticipated that the scaling behavior of a bound hole captured by a negatively charged defect will deviate from that of a typical Rydberg exciton, owing to the difference in linewidth broadening mechanisms as discussed above. It should be emphasized that the estimation of the inverse power law on $n$ in Fig.~\ref{fig:linewidthdependence} is rather imprecise due to limited data for higher excited states and simplifications in the scattering rate model, indicating a need for further investigations to determine the precise dependence of the linewidths on the principal quantum number.

\section{Capture cross-sections}
Typically, the nonradiative charge carrier capture coefficient $C$ is represented in terms of capture cross-sections $\sigma_\text{cap}$ and charge carrier's thermal velocity $\nu$ where
\begin{equation}
C=\nu \sigma_\text{cap}
\end{equation}
Alkauskas et al. \cite{Alkauskas2014} proposed a computationally convenient quantity to calculate which relates the charge carrier capture coefficient via
\begin{equation}
C=V\Gamma_\text{cap}
\end{equation}
where $V$ is the volume of the system and $\Gamma_\text{cap}$ is the capture rate of the charge carriers. Effectively, the capture cross-section is then
\begin{equation}
\sigma_\text{cap}=\frac{V\Gamma_\text{cap}}{\nu}
\end{equation}
Under the assumption of thermal equilibrium (TE), the capture rate is approximated as the total of the direct emission rates from all of the excited levels to the GS, weighted by the partition functions. In our case, the assumption of TE implies the inclusion of both bound and free hole states in the whole diamond. Consequently, the total partition function $Z_\text{tot}$ is the sum of the partition functions due to the bound states $Z_\text{B}$ and the free hole states $Z_\text{F}$. Accordingly, the capture rate is given by
\begin{equation}
\Gamma_\text{cap}=\sum_{b,i_\text{B},i_\text{F}} \frac{1}{Z_\text{tot}}\Bigg[\left(e^{-\beta\Delta E_{b,i_\text{B}}}\Gamma_{0}^{b,i_\text{B}}+e^{-\beta\Delta E_{b,i_\text{F}}}\Gamma_{0}^{b,i_\text{F}}\right)\Bigg]
\end{equation}
where 
\begin{equation}
Z_\text{tot}=\sum_{b,i_\text{B},i_\text{F}}e^{-\beta\Delta E_{b,i_\text{B}}}+e^{-\beta\Delta E_{b,i_\text{F}}}
\end{equation}
Here, $\Delta E_{b,i_\text{B}}$ is the energy difference between the $i^{th}$ bound state and $E_1$ where $E_1$ is the energy of the first ES. $\beta=1/k_BT$ where $k_B$ is the Boltzmann constant and $T$ is the temperature. $\Gamma_0^{b,i_\text{B}}$ and $\Gamma_0^{b,i_\text{F}}$ denotes the direct emission rates from the bound states and the free states to the GS, respectively, permitted by the phonon energies. Under the one-phonon scattering model, there are no direct transitions from the free hole states to the GS due to the absence of sufficiently high energy phonons and therefore $\Gamma_0^{b,i_\text{F}}\approx 0$. However, the free hole states do contribute to the partition function.

\par For the partition function of the free hole state, we convert the discrete sum of wavevectors into a continuous variable where
\begin{equation}
Z_\text{F}=\frac{V}{\left(2\pi\right)^3}\int_{0}^{\infty}\int_{0}^{\pi}\int_{0}^{2\pi} e^{-\beta\epsilon}k^2\sin\theta dk d\theta d\phi 
\end{equation}
where $\epsilon =-E_1+\kappa k^2$ is the energy of the free hole states with constant $\kappa$ determined by the curvature of the VBM in the band structure. The resulting integral is
\begin{equation}
Z_\text{F}=\frac{\epsilon^{\beta E_1}V}{8\left(\pi \kappa \beta\right)^{3/2}}
\end{equation}
We calculate the curvature $\kappa$ by averaging over the contributions from the HH and LH bands where
\begin{equation}
\kappa^{3/2}=\frac{2}{3}\left(\frac{\hbar^2}{2m_\text{HH}}\right)^{3/2}+\frac{1}{3}\left(\frac{\hbar^2}{2m_\text{LH}}\right)^{3/2}
\end{equation}
\par To assess the contributions of $Z_\text{B}$ at large $V$, we adopt the partition function of a hydrogen atom for the bound hole state \cite{Ebeling2012} where
\begin{equation}
Z_\text{B}\approx \sum_{n=1}^{\infty} n^2 \left(e^{-\beta E_n}-1+\beta E_n\right)
\label{eq:Hpartitionfunction}
\end{equation}
$E_n=R\left(1-1/n^2\right)$ where $R=13.6~\text{eV}$ and we have taken the GS as a reference level for the energy. We terminate the sum in Eq.~\ref{eq:Hpartitionfunction} over the bound states by the maximum principal quantum number $n$ that corresponds to the wavefunctions which fits within $V$. To estimate the value of $n$, we approximate the size of the electronic orbital by its Bohr's radius $r=n^2a_0$. Since Eq.~\ref{eq:Hpartitionfunction} scales with $n^2$ and thus it scales with $r$, therefore we find that $Z_\text{B}$ effectively scales as $V^{1/3}$. In the limit of large $V$, the contribution of $Z_B$ is therefore negligible compared to $Z_\text{F}$. Consequently, using an estimated hole thermal velocity of $\nu\approx 1.1\times 10^5~\text{ms}^{-1}$, the capture cross-section which is valid at room temperature is approximately
\begin{equation}
\sigma_\text{cap}\approx \frac{8}{\nu}e^{-\beta E_1}\left(\pi\kappa\beta\right)^{3/2}  \sum_{b,i_\text{B}}e^{-\beta\Delta E_{b,i_\text{B}}}\Gamma_0^{b,i_\text{B}}
\label{eq:capturecs}
\end{equation}

\section{Photoionization spectrum of NV$^0$ $\rightarrow$ NV$^-$ + bound hole}
Accurate modeling of the phonon sideband (PSB) for NV$^0$ $\rightarrow$ NV$^-$ + bound hole using \textit{ab initio} techniques is impossible due to computational constraints. In this section, we construct the photoionization spectrum of NV$^0$ $\rightarrow$ NV$^-$ + bound hole by adopting the Huang-Rhys model to describe the zero-phonon lines (ZPL) and the PSBs \cite{Stoneham2001}. We first derive an expression for the transition dipole moment matrix elements, which we then evaluate using \textit{ab initio} techniques. Next, we construct the vibrational overlap functions to estimate the NV$^0$ $\rightarrow$ NV$^-$ + bound hole PSBs. Finally, we evaluate the absorption cross-section to obtain the final photoionization spectrum.

\subsection{Transition dipole moments}
In the single particle state, we can write the transition dipole moment matrix elements $\vec{D}_{b,i}$ as
\begin{equation}
\vec{D}_{b,i}=\left< e \left| \vec{r}\right|\psi_{b,i}\right> 
\end{equation} 
where we have explicitly expressed the state $\psi_{b,i}$ in terms of the $i^{th}$ energy level of the $b^{th}$ band at the VBM. Expressing the above equation in terms of the overlap integral, the transition matrix element is 
\begin{equation}
\vec{D}_{b,i}=\int_{-\infty}^{\infty} e^*\left(\vec{r}\right) \vec{r}F_{b,i}\left(\vec{r}\right)u_{b}\left(\vec{r}\right) d^3 r
\end{equation}
where $F_{b,i}$ are the envelope wavefunctions obtained from the bound state simulations in Sec.~\ref{sec:simulateboundhole}. The observed transition is the transition of a hole from the $e$ orbital of NV$^0$ to the bound state. Assuming the $e$ orbital of NV$^0$ is sufficiently localized that it is only non-zero over the dimension of the NV$^-$ center, then we may perform Taylor series expansion and express the transition matrix element as 
\begin{equation}
\vec{D}_{b,i}=F_{b,i}\left(\vec{R}_{NV}\right)\underbrace{\int_{-\infty}^{\infty} e^*\left(\vec{r}\right)\vec{r}u_{b}\left(\vec{r}\right) d^3 r}_{\vec{d}_{b}} +\frac{\partial F_{b,i}}{\partial \vec{r}}\Bigg|_{\vec{R}_{NV}}\cdot \underbrace{\int_{-\infty}^{\infty} e^*\left(\vec{r}\right) \vec{r}\vec{r}u_{b}\left(\vec{r}\right) d^3 r}_{\overleftrightarrow{Q}_{b}}
\label{eq:transitionmatrixelements}
\end{equation}
where $\vec{d}_{b}$ and $\overleftrightarrow{Q}_{b}$ are the dipole and the quadrupole moments, respectively, which can be calculated using \textit{ab initio} techniques. The zero-order term corresponds to the amplitude of the envelope wavefunction at the NV$^-$ center while the first-order term is the gradient. For p-like transitions, only the first-order term is non-zero.

\subsection{\textit{Ab initio} calculations}
\label{sec:sidebandDFT}

\begin{figure}[h]
\centering
\includegraphics[width=0.63\linewidth]{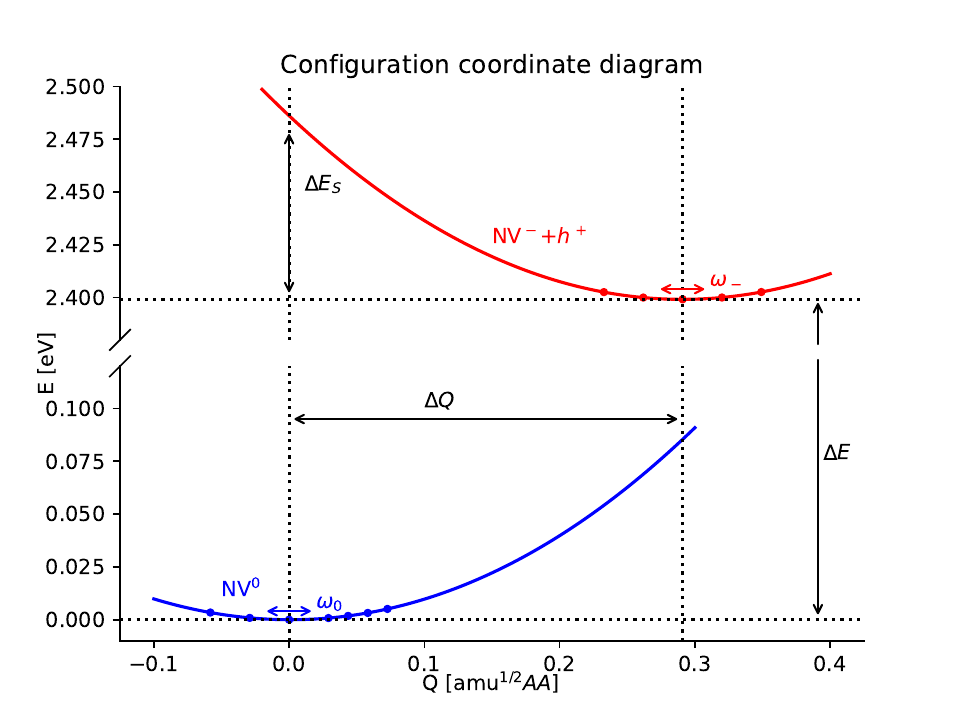}
\caption{One-dimensional configuration coordinate diagram of the NV center for a hole capture process: NV$^0$ $\rightarrow$ NV$^- +h^+$. The free hole is denoted as $h^+$. The blue surface shows the NV$^{0}$ in the $^2$E ground state and red shows NV$^{-}$ in $^3A_2$ the ground state with $h^{+}$. $\Delta Q$ is the mass-weighted atomic displacement between the minima of the two different charge states, $\Delta E_s$ denotes the Stokes shift of NV$^{-}$ and $\Delta E$ denotes the energy difference between the ground state of the two charge states. We also show the harmonic frequencies $\omega_0$ and $\omega_-$ that describe the shown surfaces of NV$^{0}$ and NV$^{-}$, harmonically. Dots denote calculated values and solid lines give the interpolated surfaces. The calculations used a $4\times4\times4$ (512)-atom supercell using the HSE06 functional.}
\label{fig:ccdiagram}
\end{figure}

\par In this work, we characterize the NV$^{0}$ and NV$^-$ center in diamond within spin-polarized DFT as implemented in the VASP code  \cite{Kresse1996a,Kresse1994,Kresse1996,Kresse1999} with HSE06 exchange-correlation functional~\cite{Krukau2006}. The $4\times4\times4$ (512 atom) supercell has been created by converging the cubic diamond unit cell containing 8 atoms with $12\times12\times12$ $k$ points and the HSE06 functional leading to a lattice constant of 3.54 \AA. Using this value, we construct the supercells including the NV center and optimize the structures to a force tolerance of $10^{-4}$~eV/\text{\AA} per ion with plane wave energy cutoff of 420 eV. We further constrain the NV$^0$ supercell to the high symmetry configuration of $C_{3v}$ to avoid the dynamic Jahn-Teller effect, which would lower the symmetry to C$_{1h}$ in the $^2$E ground state \cite{Zhang2018} for NV$^0$.

\begin{figure}[h!]
\centering
\includegraphics[width=0.5\linewidth]{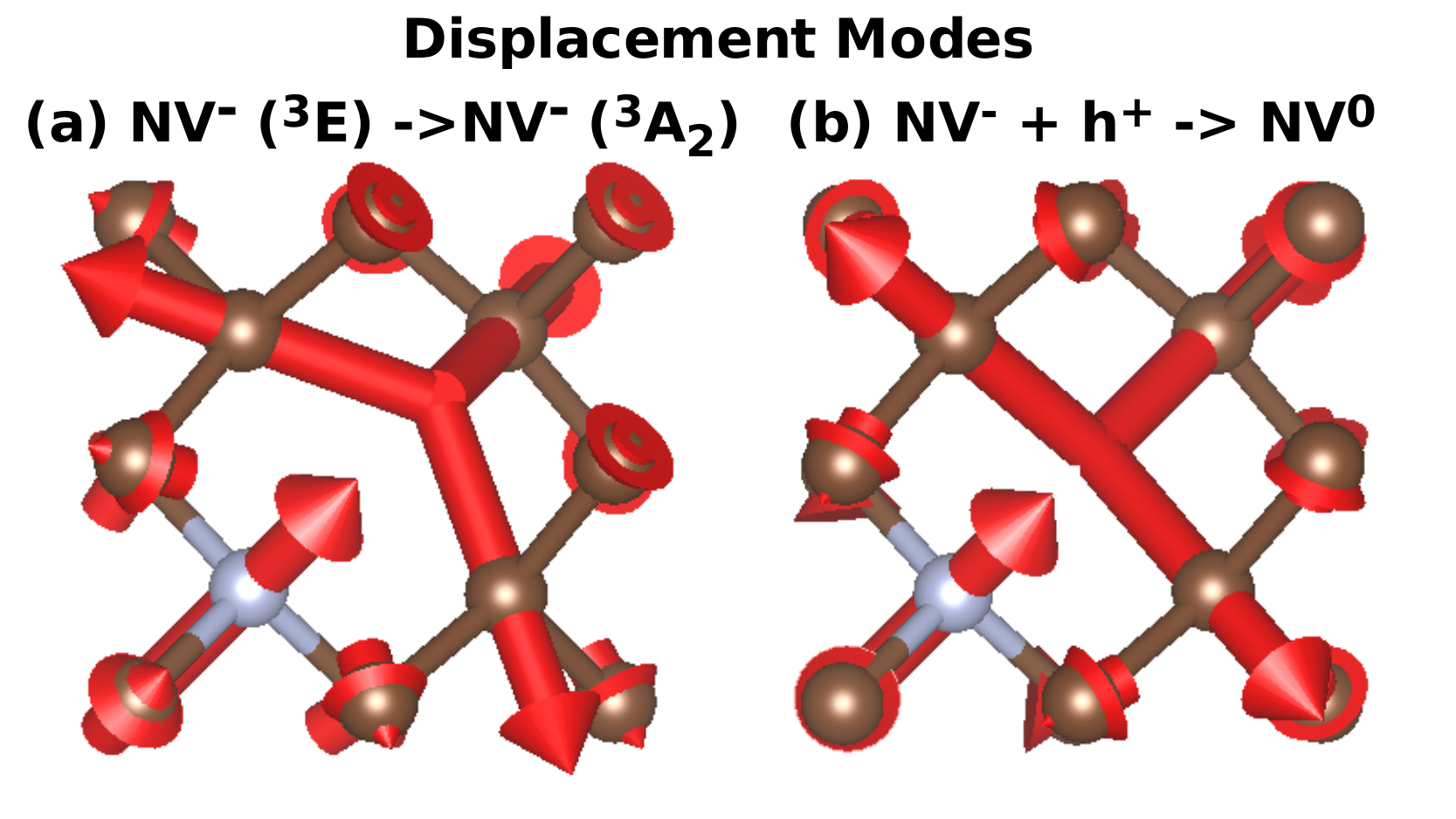}
\caption{Illustration of the displacement vectors of the internal transition in NV$^-$~\cite{Alkauskas2014} and the $\text{NV}^0\rightarrow \text{NV}^- + h^+$ transition. Both vectors have an overlap of 71.3\%.} 
\label{fig:displacementmodes}
\end{figure}

\begin{table}[h!]
\caption{Values defined in the configuration coordinate diagram of the NV center for a hole capture process: $\text{NV}^0 \rightarrow \text{NV}^- + h^+ $ and the  NV$^-$ ($^3A_2$) $\rightarrow$ NV$^-$ ($^3E$) transition. Latter values are taken from Ref.~\cite{Alkauskas2014}}
\label{tab:ccdiagram}
\begin{ruledtabular}
\begin{tabular}{ccc}
Quantity & $\text{NV}^0 \rightarrow \text{NV}^- + h^+ $   & $\text{NV}^-\left(^3A_2\right)\rightarrow \text{NV}^- \left(^3E\right)$ \cite{Alkauskas2014} \\ 
 \hline
 $\Delta E$ & 2.399 eV & 1.945 eV (experimental)\\ 
 $\Delta Q$ & 0.291 amu$^{1/2}$\AA & 0.71  amu$^{1/2}$\AA\\ 
 $\hbar\omega_0$ & 90.93~meV& - \\
 $\hbar\omega_-$ & 92.75~meV&~65meV\\
 $\Delta E_S$ & 87.41~meV& -\\ 
\end{tabular}
\end{ruledtabular}
\end{table}

\par
\begin{figure}[h!]
\centering
\includegraphics[width=0.8\linewidth]{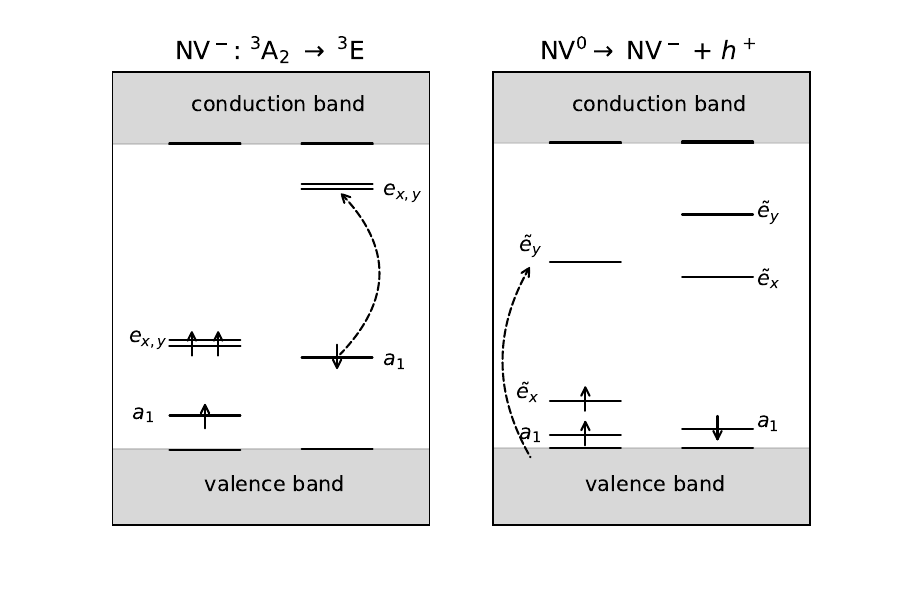}
\caption{Schematics of the excitation for the $\text{NV}^- \left(^3A_2\right) \rightarrow \text{NV}^- \left(^3E\right)$ (left) and the $\text{NV}^0\rightarrow \text{NV}^- + h^+$ (right). In contrast to the transitions in the NV$^-$ center, which are between two well localized orbitals, the transitions of the NV$^0$ center are between localized and more delocalized orbitals.}
\label{fig:transitionschematics}
\end{figure}

 In Fig.~\ref{fig:displacementmodes}, we show the normalized displacement vectors $\Delta Q$ for the $\text{NV}^0 \rightarrow \text{NV}^- + h^+ $ as discussed in Fig.~\ref{fig:ccdiagram} and the internal transition of $\text{NV}^-\left(^3A_2\right)\rightarrow \text{NV}^- \left(^3E\right)$ \cite{Alkauskas2014} that leads to the ZPL line in the spectrum of the NV$^{-}$ center. Fig.~\ref{fig:transitionschematics} shows the schematics of these two different transitions. Overall, the overlap between these two vectors is 71.3\%. We find that the displacement on the N atom exhibits greater similarity between the two displacement modes, in contrast to the displacements on the neighboring C atoms. Additionally, from Tab.~\ref{tab:ccdiagram}, we find that the displacement vector in the  $\text{NV}^-\left(^3A_2\right)\rightarrow \text{NV}^- \left(^3E\right)$ has a magnitude which is 2.4 times larger, while the harmonic phonon frequency is smaller than in the $\text{NV}^0 \rightarrow \text{NV}^- + h^+ $ transition. 

\par Next, we show the transition dipole moments $\vec{d}$ for the $\text{NV}^0 \rightarrow \text{NV}^- + h^+ $ excitation with a comparison to the $\text{NV}^-\left(^3A_2\right)\rightarrow \text{NV}^- \left(^3E\right)$ in Tab.~\ref{tab:transitiondipoleboundhole} and Tab.~
\ref{tab:transitiondipoleinternal}, respectively. Due to the delocalized nature of the excitation, we have to calculate large supercells and rely on an extrapolation scheme. In the case of $\text{NV}^0 \rightarrow \text{NV}^- + h^+ $, we use the nuclear and electronic configuration of the NV$^0$ ground state to calculate the transition dipole moment between the Kohn-Sham states. The presence of an NV center reduces the cubic symmetry of diamond to $C_{3v}$ symmetry. Consequently, this splits the triply degenerate states at the $\Gamma$ point of the VBM to a pair of doubly degenerate states and one single degenerate state. We identified two degenerate transitions from the valence band states to the $e_y$ orbital of NV$^0$, which corresponds to HH. However, LH transition could not be identified. For the $\text{NV}^-\left(^3A_2\right)\rightarrow \text{NV}^- \left(^3E\right)$ transition dipole moments $\vec{\mu}$, we use the relaxed cell of the ground state of NV$^-$. We further average over the two (quasi) degenerate states, respectively. The calculation of transition dipole moments involves not only the use of a 512-atom supercell but also extends to employing up to 8000 atoms for PBE and up to 1000 atoms for HSE06. (see Tab.~\ref{tab:transitiondipoleinternal} and Tab.~\ref{tab:transitiondipoleboundhole}.)

\begin{table}[h!]
\caption{Values of the transverse transition dipole moments for the $\text{NV}^- \left(^3A_2\right) \rightarrow \text{NV}^- \left(^3E\right)$ transitions. The first column contains the number of atoms in the supercell. The transversal transition dipole moment $\left|\vec{\mu}_\perp\right|$  and total transition dipole moment $|\vec{\mu}|^2$ are given in units of Debye and Debye$^2$, respectively. The longitudinal transition dipole moment $|\vec{\mu}_\parallel|^2$ is zero.}
\label{tab:transitiondipoleinternal}
\begin{ruledtabular}
\begin{tabular}{ccccc} 
$N$ &$|\vec{\mu}_{\text{PBE},\perp}|$  &$|\vec{\mu}_\text{PBE}|^2$   & $|\vec{\mu}_{\text{HSE},\perp}|$  &$|\vec{\mu}_\text{HSE}|^2$ \\
 \hline
64  &3.6899 &13.6154    &2.9053 &8.4409\\
216 &5.1198 &26.2127    &4.3986 &19.3477\\
512 &5.6020 &31.3829    &4.8050 &23.0881\\
1000    &5.7237 &32.7613    &4.8959 &23.9694\\
1728&   5.7545  &33.1143    &   &\\
2744    &5.7621 &33.2016    &   &\\
4096    &5.7654 &33.2398    &   &\\
5832    &5.7675 &33.2638    &   &\\
8000    &5.7677 &33.2663    &   &\\\hline
 \end{tabular}
\end{ruledtabular}
\end{table}

\begin{table}
\caption{Values of the transition dipole moments for the $\text{NV}^{0} \rightarrow  \text{NV}^{-} + h^{+}$ transition. The first column contains the number of atoms in the supercell. The longitudinal/transversal transition dipole moment $|\vec{d}_\text{$\parallel/\perp$}|$  and total transition dipole moment $|\vec{d}|^2$ are given in units of Debye and Debye$^2$, respectively. The two rows in each supercell simulations correspond to possible transitions from the doubly degenerate valence band states to the $e_y$ orbitals.}
\label{tab:transitiondipoleboundhole}
\begin{ruledtabular}
\begin{tabular}{ccccccc} 
$N$ &$|\vec{d}_\text{PBE,$\parallel$}|$  &$|\vec{d}_\text{PBE,$\perp$}|$  & $|\vec{d}_\text{PBE}|^2$ & $|\vec{d}_\text{HSE,$\parallel$}|$  &$|\vec{d}_\text{HSE,$\perp$}|$  & $|\vec{d}_\text{HSE}|^2$\\
 \hline
64  & 0.0000 & 1.3035 & 1.6991 & -0.0019 & 1.8664  & 3.4834\\
    & 1.2231 & -2.1025 & 5.9165 & -1.1738 & 1.3655 & 3.2423\\\hline
216 & -0.0001 & -0.5976 & 0.3571 & 0.0000 & -0.0194 & 0.0004 \\ 
    & -0.8666 & -1.2634 & 2.3472 & -0.4492 & -0.8003 & 0.8423 \\ \hline
512 & 0.0009 & -0.3355 & 0.1125 & 0.0000 & 0.0483 & 0.0023 \\ 
    & -0.5737 & 0.8007 & 0.9702 & 0.2867 & 0.5142 & 0.3466 \\ \hline 
1000 & 0.0000 & 0.2227 & 0.0496 & 0.0000 & 0.1007 & 0.0101 \\ 
     & 0.4025 & 0.5537 & 0.4686 & 0.2080 & 0.3526 & 0.1676 \\ \hline 
1728 & 0.0003 & -0.1616 & 0.0261 & & & \\ 
     & 0.2992 & -0.4076 & 0.2557 & & & \\ \hline 
2744 & 0.0025 & -0.1899 & 0.0361 & & & \\ 
     & -0.2562 & -0.3262 & 0.1721 & & & \\ \hline 
4096 & 0.0003 & 0.0991 & 0.0098 & & & \\ 
     & 0.1869 & -0.2517 & 0.0983 & & & \\ \hline 
5832 & -0.0665 & 0.0470 & 0.0066 & & & \\ 
     & 0.0461 & 0.2541 & 0.0067 & & & \\ \hline 
8000 & -0.0002 & 0.1029 & 0.0106 & & & \\ 
     & -0.1426 & 0.1803 & 0.0528 & & & 
 \end{tabular}
 \end{ruledtabular}
\end{table}

\clearpage 

\subsection{Phonon sidebands of NV$^0$ $\rightarrow$ NV$^-$ + bound hole}
\label{sec:PSBs}
From Sec.~\ref{sec:sidebandDFT}, the displacement vectors of the internal transitions of NV$^-$ and the photoionization transition of NV$^0$ agree to 71\%. Consequently, as a first attempt, we construct the PSBs of NV$^0$ $\rightarrow$ NV$^-$ + bound holes by using the known generating function $p\left(\omega\right)$ of the NV$^-$ PSBs \cite{Kehayias2013}. The PSBs are obtained by calculating the temperature-dependent vibrational overlap function $P\left(\omega,T\right)$ using the procedures as described in Ref.~\cite{Davies1974}. 

\par
\begin{figure}[h!]
\centering
    \begin{subfigure}[t]{0.7\textwidth}
    \includegraphics[width=\textwidth]{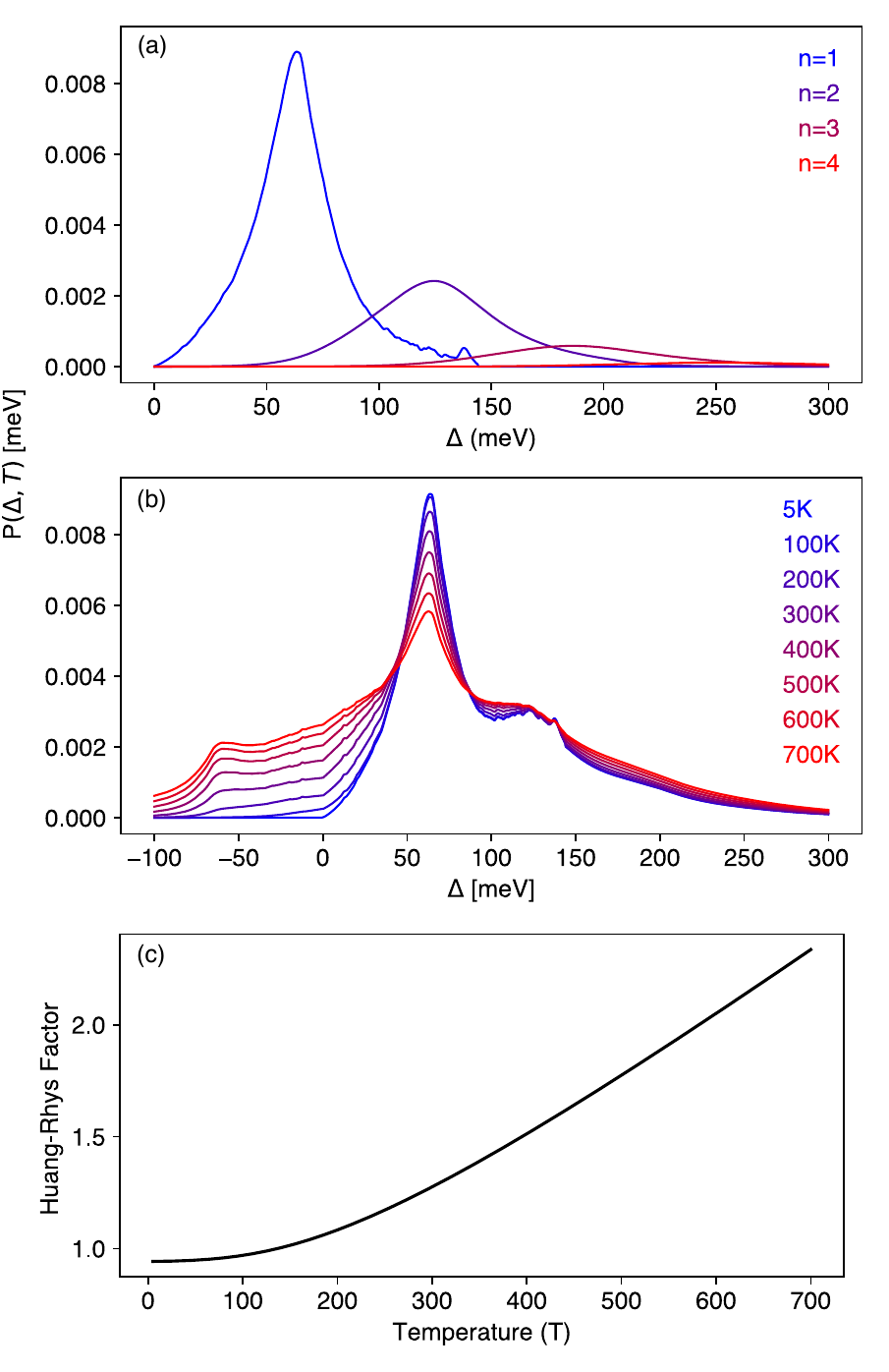}
    \phantomcaption
    \label{fig:PSBa}
    \end{subfigure}
    \begin{subfigure}[t]{0\textwidth}
    \includegraphics[width=\textwidth]{PSBs.pdf}
    \phantomcaption
    \label{fig:PSBb}
    \end{subfigure}
    \begin{subfigure}[t]{0\textwidth}
    \includegraphics[width=\textwidth]{PSBs.pdf}
    \phantomcaption
    \label{fig:PSBc}
    \end{subfigure}
\caption{(a) The approximated $n$-phonon components of the NV$^0$ optical emission PSB at $T=5~$K (increasing from $n=1$ from left to right). The sum of these components equals to the calculated $T=5~$K PSB depicted in (b). (b) The calculated temperature-dependent vibrational overlap functions $P\left(\omega,T\right)$. (c) Huang-Rhys factor $S$ plotted as a function of temperature.}
\label{fig:PSBs}
\end{figure}

 The temperature-dependent one-phonon vibrational overlap function is 
\begin{equation}
P_1\left(\omega,T\right)=
\begin{cases}
\left[n_B\left(\omega,T\right)+1\right]p\left(\omega\right) \text{ if } \omega\geq 0 \\ 
n_B\left(\omega,T\right)p\left(-\omega\right)\ \ \ \ \text{ if } \omega < 0 \\ 
\end{cases}
\end{equation}
We then calculate the temperature-dependent multiphonon vibrational overlap functions recursively by using
\begin{eqnarray}
P_i\left(\omega,T\right)&=&P_{i-1}\left(\omega,T\right)\otimes P_1\left(\omega,T\right)\nonumber \\ 
&=&\int_{-\infty}^{\infty} P_{i-1}\left(\omega-\omega',T\right)P_1\left(\omega',T\right) d\omega'
\end{eqnarray}
where $\otimes$ denotes the operation of convolution. The total temperature-dependent vibrational overlap function is given by
\begin{equation}
P\left(\omega,T\right)=e^{-S}\sum_{i=1}^{\infty} \frac{S^i}{i!}P_i\left(\omega,T\right)
\end{equation}
Here, $S$ is the temperature-dependent Huang-Rhys factor given by
\begin{equation}
S=S_0 \int_0^{\Omega} \left[2n_B\left(\omega',T\right)+1\right]p\left(\omega'\right)d\omega'
\end{equation}
where $S_0=\Delta E_\text{Stokes}/\hbar\omega_-\approx0.942$ is the low-temperature Huang-Rhys factor of the effective local phonon modes of NV$^0$ and $\Omega \approx 165~\text{meV}$ is the maximum phonon energy in diamond. We reconstruct the $n$-phonon components of the NV$^0$ optical emission PSB at $T=5~K$ as shown in Fig.~\ref{fig:PSBa} whereas Fig.~\ref{fig:PSBb} shows the calculated total temperature-dependent vibrational overlap functions $P\left(\omega ,T\right)$. The temperature-dependent Huang-Rhys factor is plotted as a function of temperature, shown in Fig.~\ref{fig:PSBc} where 0.942 and 1.275 is the Huang-Rhys factor for $T=5~$K and $T=300~$K, respectively.

\clearpage 

\subsection{Absorption cross-section}
Having constructed the PSBs in Sec.~\ref{sec:PSBs}, as an example, we now construct the absorption cross-section consisting of ZPL and PSBs, summed over all possible transitions at energy $E_{b,i}$ where
\begin{equation}
\sigma\left(E\right)= 4\pi^2\chi\sum_{b,i} E \overline{\left|\vec{D}_{b,i}\right|}^2 I_{b,i}\left(E\right)
\end{equation}
in which
\begin{equation}
I_{b,i}\left(E\right)=L\left(E-E_{b,i},\Gamma_{b,i}\right)+P\left(E-E_{b,i},T\right)
\end{equation}
As mentioned earlier, due to the inability to determine the transition dipole moments for LH transitions, we constructed the final transition dipole moments matrix elements by averaging $|\vec{d}_{\text{HSE},\parallel}|^2$ of $N=1000$ over the two possible transitions as indicated in Tab.~\ref{tab:transitiondipoleboundhole}. This averaging process resulted in a parallel-to-transverse component ratio of 0.32. Consequently, the absorption cross-section is independent of polarization. Thus, we have $\overline{\left|\vec{D}_{b,i}\right|}^2$, where the overbar denotes the averaged components. As we only obtain the transition dipole moments, our analysis are restricted to transitions that involves s-like orbitals. 
 $\chi=e^2/4\pi\epsilon\hbar c$ denotes the fine structure constant of diamond, $L$ is the ZPL of each transition modelled after a Lorentzian function (as discussed in the main text) and $P$ is the vibrational overlap functions. In explicit form, the Lorentzian function is given by
\begin{equation}
L\left(E-E_{b,i},\Gamma_{b,i}\right)= \frac{e^{-S}}{\pi} \frac{\frac{1}{2}\Gamma_{b,i}}{\left(E-E_{b,i}\right)^2+\left(\frac{1}{2}\Gamma_{b,i}\right)^2}
\end{equation}
 Taking into account the Stokes shift, the energy of NV$^-$ + bound hole state is effectively 
\begin{equation}
E_{b,i}=\underbrace{E_{\text{VBM}}}_{2940~\text{meV}} +\Delta E_{b,i}-\underbrace{E_{\text{Stokes}}}_{87\ \text{meV}}
\end{equation}

\section{Modeling other deep defects in semiconductors}
\par Reviewing the methodologies presented in this work, we summarize the key requirements to apply the methods to other deep defects in semiconductors. First, the construction of charge densities using \textit{ab initio} techniques and the resulting effective potential can only be extended to other deep defects in semiconductors if the charge densities are sufficiently localized in the defect. Second, the effective mass model and the construction of the effective potential can be applied to different materials only if the effective mass tensor is known. Third, to employ spherical symmetry approximations for defects with lower symmetry, it is necessary to conduct comparisons between the one-dimensional and three-dimensional solutions. The validity and accuracy of the spherical approximation are contingent upon small observed differences between the solutions, usually on the order of meV, or, in other words, when the errors are less than 1\%. Fourth, to apply our formalism of the deformation potential model of hole-phonon scattering, knowledge of the density, longitudinal acoustic velocity, phonon dispersion relations, and the deformation potential of the acoustic and optical phonon is required. Fifth, our capture cross-section expression is only valid at room temperature, and the thermalization rate is a critical requirement. Specifically, to assume thermal equilibrium, the emission rate of the first excited state must be slower than its absorption rate to the second excited state, which is temperature dependent. Finally, constructing the photoionization spectrum necessitates a comprehensive understanding of the PSBs of the internal transitions of the defect. The PSBs of the photoionization transitions can only be constructed if they can be well described by the PSBs of the internal transitions of the defect. Notably, the correspondence in the phonon modes associated with the transitions reflect the similarity of the phonon sidebands.

\bibliography{/Users/yunhengchen/Library/CloudStorage/OneDrive-AustralianNationalUniversity/ANU/Research/Bibliographies/mainbib.bib}